\documentclass[reprint,amsmath,amssymb,pra,aps, nofootinbib,superscriptaddress]{revtex4-1}

\usepackage[T1]{fontenc}
\usepackage[utf8]{inputenc}
\usepackage{amsmath, amssymb}
\usepackage{bbold}
\usepackage{bm}
\usepackage{tipa}
\usepackage{blindtext}
\usepackage[dvipsnames]{xcolor}
\definecolor{dcyan}{RGB}{0,100,100}
\usepackage{graphicx}
\usepackage{tabularx}
\usepackage{textgreek}
\usepackage{xfrac}
\usepackage{units}
\usepackage{comment}
\usepackage{float}

\usepackage[sectionbib]{bibunits}
\defaultbibliographystyle{naturemag}
\defaultbibliography{references}

\usepackage{hyperref}
\usepackage{xcolor}
\definecolor{green_cust}{RGB}{0,154,85}
\definecolor{red_cust}{RGB}{173,49,54}
\definecolor{blue_cust}{RGB}{0,103,148}
\hypersetup{colorlinks=true, linkcolor=red_cust, citecolor=green_cust, filecolor=blue_cust, urlcolor=blue_cust}

\usepackage{soul}

\newcommand{\Figref}[1]{Fig.~\hyperref[#1]{\ref{#1}}}

\begin{document}
\title{Quantum-controlled synthetic materials}
\author{Andrei Vrajitoarea}
\affiliation{Center for Quantum Information Physics, Department of Physics, New York University, New York 10003, USA}
\author{Gabrielle Roberts}
\affiliation{The Department of Physics, The James Franck Institute, and The Pritzker School of Molecular Engineering, The University of Chicago, Chicago, IL}
\author{Kaden R. A. Hazzard}
\affiliation{Department of Physics and Astronomy, Rice University, Houston, TX, USA}
\affiliation{Smalley-Curl Institute, Rice University, Houston, TX 77005, USA}
\author{Jonathan Simon}
\affiliation{The Department of Physics, Stanford University, Stanford, CA}
\affiliation{The Department of Applied Physics, Stanford University, Stanford, CA}
\author{David I. Schuster}
\affiliation{The Department of Applied Physics, Stanford University, Stanford, CA}
\date{\today}


\begin{abstract}
Analog quantum simulators and digital quantum computers are two distinct paradigms driving near-term applications in modern quantum science, from probing many-body phenomena to identifying computational advantage over classical systems. A transformative opportunity on the horizon is merging the high-fidelity many-body evolution in analog simulators with the robust control and measurement of digital machines. Such a hybrid platform would unlock new capabilities in state preparation, characterization and dynamical control. 
Here, we embed digital quantum control in the analog evolution of a synthetic quantum material by entangling the lattice potential landscape of a Bose-Hubbard circuit with an ancilla qubit. This Hamiltonian-level control induces dynamics under a superposition of different lattice configurations and guides the many-body system to novel strongly-correlated states where different phases of matter coexist -- ordering photons into superpositions of solid and fluid eigenstates. 
Leveraging hybrid control modalities, we adiabatically introduce disorder to localize the photons into an entangled cat state and enhance its coherence using a many-body echo technique.
This work illustrates the potential for entangling quantum computers with quantum matter -- synthetic and solid-state -- for advantage in sensing and materials characterization.
\end{abstract}

\maketitle

\section{Introduction}
\label{sec:intro}
Advances in programmable quantum simulation platforms have enabled the rapid development of synthetic matter assembled from individual quantized systems with tailored interactions between them.
Such platforms offer unique possibilities for exploring many-body physics in a highly controllable way~\cite{bloch2012quantum, blatt2012quantum, Aspuru-Guzik2012quantum, browaeys2020many, carusotto2020photonic, monroe2021programmable}, by leveraging microscopic and dynamical probes~\cite{bakr2009quantum, cheuk2015quantum} and navigating extreme parameter regimes~\cite{simon2011quantum, kollar2019hyperbolic, VrajitoareaUSC} beyond those accessible in solid-state materials. Investigating and utilizing the complex coherent dynamics in these designer many-body systems has broad implications across quantum science, from understanding fundamental phenomena~\cite{cirac2012goals}, to engineering quantum information processors robust to errors~\cite{bluvstein2022quantum, google_willow2025} and benchmarking state-of-the-art computational tools~\cite{eisert2015quantum, daley2022practical, trivedi2024quantum}.

Recent efforts in the precise manipulation of isolated programmable many-body systems have shed light on various out-of-equilibrium phenomena, including information scrambling~\cite{landsman2019verified, googlescramble2021} and thermalization~\cite{Rigol2008, Kaufmanscience2016, andersen2025thermalization}, formation and scaling of entanglement~\cite{joshi2023exploring, karamlou2024probing}, time crystallinity~\cite{zhang2017observation, choi2017observation}, many-body localization~\cite{Schreiber2015, choi2016exploring, roushan2017spectroscopic, guo2021observation}, quantum scarring~\cite{Bernien2017, bluvstein2021controlling} and anomalous transport~\cite{brown2019bad, Joshi2022, VrajitoareaUSC, BelyanskyPRR2021}.
Investigating the late-time dynamics in these analog quantum simulators offers a first practical advantage, where moderately-sized quantum systems already challenge state-of-the-art numerical methods~\cite{daley2022practical, trivedi2024quantum, andersen2025thermalization, haghshenas2025digital}.
Controlling the many-body dynamics in such programmable systems can also enable other practical applications from generating large-scale entangled states as highly-sensitive metrological probes~\cite{Degen_RMP2017, Zoller_PRL2024}, to solving optimization problems encoded in spin models~\cite{cerezo_NatRev2021, Bharti_RMP2022}.

While these synthetic materials are fundamentally quantum mechanical, \emph{all} current techniques guide their dynamics through \emph{classical} control of the system Hamiltonian, because the Hamiltonian parameters that dictate the unitary evolution are defined with external, \emph{classical} fields. In analogy to a transistor, where carrier transport across its channel is controlled by a gate voltage, transport dynamics across the system is configured with classical control signals: lattices of (classical) laser beams confine and manipulate atomic simulators; (classical) currents generate the magnetic fields that tune the frequencies of transmons. Can we envision realizing a \textit{quantum-controlled transistor}, such that the evolution of a synthetic many-body system is controlled by \emph{another} quantum system?
This capability would open the door to hybrid devices interfacing digital quantum computers with analog simulators and sensors~\cite{daley2022practical}.

Controlling \& entangling quantum matter with small quantum computers would enable novel protocols for preparing and probing strongly correlated states of matter and light, and new routes to quantum advantage~\cite{gilyen2019quantum, martyn2021grand, garratt2407quantum}. 
Similar ideas are under active exploration in sub-wavelength atomic arrays, where the state of a single ancilla atom controls the macroscopic scattering response of an array~\cite{srakaew2023subwavelength}, providing a quantum optical meta-surface that can generate highly correlated light such as Greenberger–Horne–Zeilinger and multidimensional cluster states~\cite{bekenstein2020quantum}.
Furthermore, entangling ancilla qubits with initial states can leverage ancilla-conditioned dynamics to probe anyon braiding phases~\cite{GoogleAnyons2021}, excitation spectra~\cite{roushan2017spectroscopic, Morvan2022-boundstate, roberts2024manybody}, and information scrambling~\cite{landsman2019verified, googlescramble2021}.

In this work, we demonstrate a quantum-controlled synthetic material in a Bose-Hubbard circuit of strongly interacting microwave photons confined to a lattice of coupled superconducting qubits. 
Our approach marries the robust site-resolved control of the lattice potential with the qubit control of each individual site to realize Hamiltonian-level quantum control of our many-body system by entangling the lattice energy landscape with an ancilla qubit.
Such control enables the system to evolve under a superposition of lattice configurations entangled with the state of the ancilla qubit, thus realizing a photonic transistor where quantum logic is embedded in the analog dynamics of a many-body system. The interfering trajectories guide the system to a first-of-its-kind quantum state where different phases of matter coexist, ordering the photons into a superposition of a solid (Mott insulator~\cite{Ma2019AuthorPhotons}) and fluid eigenstate~\cite{AdbPaper} entangled with the ancilla.
Following the preparation of this unconventional state,  adiabatic site-resolved control of disorder re-localizes the photons to a highly-entangled N00N (cat) state.
By entangling the lattice geometry with the state of an ancilla qubit, we can perform Ramsey measurements~\cite{roberts2024manybody} on the ancilla to learn about the impact of the lattice geometry on the many-body states. Leveraging the hybrid analog-digital control, we introduce a many-body echo protocol to enhance the coherence of the entangled states by decoupling their evolution from  low-frequency phase noise.


In what follows, we introduce our circuit platform and the ancilla-conditioned evolution protocol. We apply this protocol to assemble solid + fluid superpositions and their corresponding cat states, employing complementary schemes that utilize the precise control of the lattice potential and fluid excitations.
Finally, we employ many-body Ramsey interferometry to probe the coherence of our long-range correlated cat states, and an echo protocol to suppress slow dephasing of the many-body states.

\begin{figure}[htp]
	\includegraphics[width=0.5\textwidth]{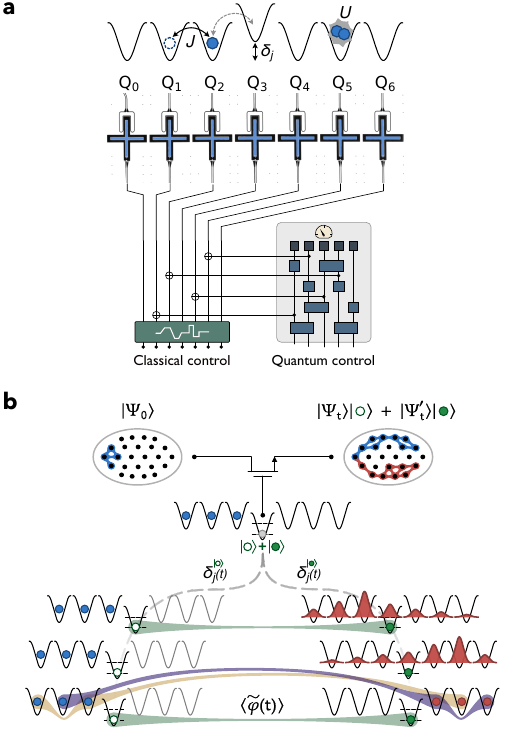}
	\caption{\textbf{Quantum control of transport in a Bose-Hubbard circuit}. \textbf{a.} The physical platform used to investigate transport dynamics maps to a 1D Bose-Hubbard model where bosonic particles coherently propagate in a tight-binding lattice and experience strong local interactions. This system is implemented as a chain of capacitively coupled superconducting transmon qubits (blue), serving as the lattice sites that host photonic particles (microwave excitations of the qubits). Inter-site capacitive coupling mediates particle tunneling ($J$), and the transmon anharmonicity provides local interactions ($U$). Site-resolved control of the lattice potential landscape is achieved by tuning the transmon resonance energies using inductively coupled bias lines to thread magnetic flux through their SQUID loops. Leveraging the quantum nature of the lattice sites, specifically their occupancy-dependent energy transition, offers a unique capability for steering the entanglement dynamics in our many-body system, by marrying the state preparation protocol of correlated fluids~\cite{AdbPaper} with the quantum control of the lattice energy landscape.
    \textbf{b.} 
    Schematic representation of a quantum-controlled photonic transistor, 
    where the many-body dynamics is regulated by the state of a control site. A superposition of zero and one particles in the control site drives interference between different lattice configurations to generate entangled states with long-range correlations, which we characterize by extracting the phase information of the control qubit.}
    \label{fig:setupfig}
\end{figure}

\section{Quantum Circuit Platform}
\label{sec:circuit&protocol}
Our experiments take place in the platform shown in Fig.~\ref{fig:setupfig}b). The dynamics of this circuit is captured by a 1D Bose-Hubbard model (Fig.~\ref{fig:setupfig}a)
\begin{align*}
\mathrm{H}_\mathrm{BH}/\hbar = J &\sum_{ \langle i,j \rangle}{a_i^\dagger a_j^{\phantom \dagger}
 } + \frac{U}{2}\sum_i{n_i \left(n_i-1\right)}\\
+ &\sum_i {[\omega_\mathrm{0}+\delta_i(t)] n_i} \label{eq:bosehubbardC},
\end{align*}

\noindent describing the coherent propagation of bosonic particles (photons) in a lattice, with a nearest-neighbor tunneling rate $J$, experiencing strong local interactions with an on-site energy $U$. The operator $a_i^\dagger$ ($a_i$) creates (destroys) a microwave photon on site $i$, where the energy of the first photon is $\omega_0 +\delta_i$, and $\hbar$ is the reduced Planck’s constant.
\newline
\indent In our superconducting circuit platform~\cite{carusotto2020photonic, AdbPaper, MBRamPaper} the tight-binding lattice sites are implemented as transmon qubits~\cite{Koch2007}, the bosonic particles are the qubit's microwave excitations, the tunneling ($J$) is achieved through the capacitive coupling of neighboring qubits, and the on-site interactions ($U$) stem from the transmon anharmonicity. 
The lattice energy landscape is dynamically controlled with site-resolved precision by individually tuning the transmon frequencies using flux bias lines (see SI~\ref{SI:FluxControlandCrosstalk}). We operate the device with the static parameters $J/2\pi = -9$~MHz, $U/2\pi = -240$~MHz, and a qubit tuning range $(\omega_0 + \delta_i)/2\pi \in [3,6]$~GHz where we investigate the lattice dynamics at $\omega_0/2\pi\approx 5.31$~GHz. The photon lifetime $T_1 \approx 45$\,\textmu s allows many-body effects to dominate over dissipation $|U|\gg |J| \gg 1/T_1$. Each transmon is capacitively coupled to an off-resonant coplanar waveguide resonator to allow site-resolved microscopy of the photon occupation (see SI~\ref{SI:SystemparametersandOperatingpoints}).
\newline
\indent Capitalizing on the site-resolved control of the lattice potential profile has played a pivotal role in preparing and characterizing many-body states in our synthetic platform. We have harnessed the adiabatic control of lattice disorder ($\delta_i \gg J$) to prepare fluids of light~\cite{AdbPaper}, and subsequently engineered lattice perturbations ($\delta_i \approx J$) in real-time ($t \ll J^{-1}$) to probe many-body dynamics~\cite{vrajitoarea_soundwaves}. These techniques have thus far relied entirely on classical control of the lattice. 
We exploit the quantum nature of the actual lattice sites (the transmon qubits), where their transition energy depends upon their occupancy, to create lattice energy landscapes that depend upon the occupancy of several sites. When incorporating state-dependent potentials with superpositions of these occupancies, the system undergoes evolution under a superposition of different lattice configurations, allowing for unconventional control of the many-body dynamics.
\newline
\indent  This capability becomes the enabling ingredient for implementing a quantum-controlled photonic transistor using ancilla-conditioned lattice potentials (Fig.~\ref{fig:setupfig}c). 
The protocol relies on frequency detuning a lattice site by its interaction energy $U$, to mediate the transport of particles through its doubly-excited state.
This resonant tunneling process is allowed \emph{only} if the lattice site is occupied with a photon, thereby making the entire dynamics of the many-body system conditioned on the quantum state of the lattice (control) site. This transistor protocol relies on injecting a superposition of zero and one particles into the control site, enabling the coherent interference of evolution trajectories across different spatial sectors of the lattice.
Introducing disorder enables us to controllably freeze the dynamics and prepare a many-body entangled state, with long-range correlations, which we efficiently characterize through Ramsey interferometry of the control qubit.

\section{Entanglement protocol}
\label{sec:superpositions}

\begin{figure*}[ht] 
	\centering
 	\includegraphics[width=1\textwidth]{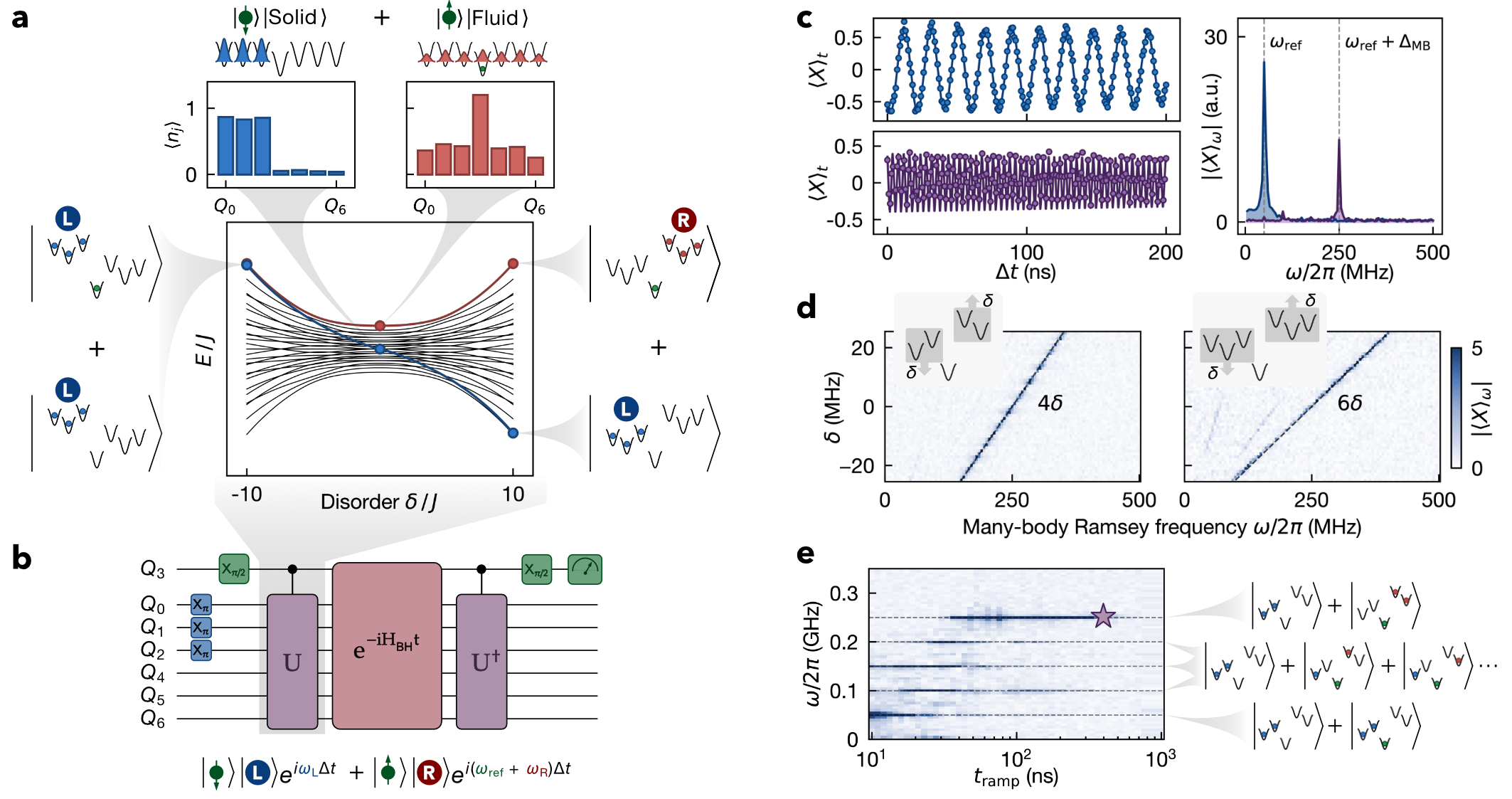}
	\caption{
		\textbf{Many-body entangling operation via ancilla-conditioned transport}. \textbf{a.} Starting in a highly disordered configuration (computational basis), we initialize the highest energy state by exciting the qubits on the left-hand side, and adiabatically remove disorder to a quantum-controlled transport configuration, where the middle site is detuned by $U$ and mediates dynamics between the two halves of the system through its doublon state. If this ancilla qubit is in its ground (excited) state, the particles localize (delocalize) into a solid (fluid), as shown by the measured density profile in blue (red). Preparing the ancilla in a particle + hole superposition ($\frac{\pi}{2}$ pulse) produces a solid + fluid superposition, which maps to a many-body entangled (N00N) state in the computational basis when we adiabatically relocalize the photons to an inverted disordered configuration. \textbf{b.} The coherence of such entangled states is probed through many-body Ramsey interferometry: we coherently evolve the entangled state for a time $\Delta t$ to accumulate a relative phase proportional to the energy difference of the left and right qubit clusters, and apply the corresponding disentangling operation to relocalize the phase information into the ancilla qubit accessible from its population measurement following a second $\frac{\pi}{2}$ pulse. \textbf{c.} The frequency difference in the ancilla interference fringe when it is uncoupled (blue) and entangled (purple) with the lattice matches the energy difference between the left and right qubit clusters $\omega_L  - \omega_R$, consistent with the preparation of the coherent cat-like superposition of these two collective states. \textbf{d.} To leverage this N00N state, we benchmark our system's capability as a sensor:  we offset the energy of the right/left qubit clusters by $\pm\delta$ during the phase evolution and observe an increased sensitivity, where the shift in the many-body Ramsey frequency scales with system size $(N-1)\delta$ for both $N=5$ (left) \& $N=7$ (right) qubit N00N states. \textbf{e.} Adiabaticity is characterized by varying the ramp time $t_\mathrm{ramp}$, where discrete changes in the Ramsey Fourier frequencies reveal the transition from a product to a fully entangled state. 
	}
	\label{fig:adbNOON}
\end{figure*}

The conditional transport protocol is presented in Fig.~\ref{fig:adbNOON}a. First, we prepare the lattice in the highest energy three-particle state: the imposed disorder localizes \emph{all} many-body eigenstates, enabling us to prepare this state by injecting individual photons into the three sites on the left half of the lattice ($Q_0, Q_1, Q_2$), via site-resolved microwave $\pi$ pulses. Additionally, a central ancilla transmon ($Q_3$) serves as the quantum switch prepared in an arbitrary single-particle state $|\varphi\rangle$.
In this disordered lattice configuration, the many-particle eigenstate corresponds to a localized product state $|\Psi_i\rangle = |111\rangle |\varphi\rangle|000\rangle$. 
We then adiabatically remove the disorder to a quantum-controlled transport configuration, where the middle site is detuned by $U$ and the evolution of the initial product state depends upon the occupancy of this middle site/ancilla qubit. If the ancilla qubit is empty $|\varphi\rangle = |0\rangle$, the particles remain in a localized state corresponding to a Mott insulator of the disorder-free system~\cite{Ma2019AuthorPhotons}, with a Mott insulator-vacuum domain wall at the center. If the ancilla qubit is excited with a photon $|\varphi\rangle = |1\rangle$, then its (now resonant) doubly excited state allows for tunneling into the other half of the lattice, resulting in a correlated fluid~\cite{AdbPaper}. The measured density profiles for these two distinct ancilla-controlled states are highlighted in the inset of Fig.~\ref{fig:adbNOON}a. 
The central result of this protocol is the production of a superposition of solid + fluid many-body states through the manipulation of the ancilla: preparing the ancilla qubit in a particle + hole superposition $|\varphi\rangle = \frac{|0\rangle + |1\rangle}{\sqrt{2}}$ using a $\frac{\pi}{2}$ pulse. The final step is to relocalize the photons by adiabatically re-introducing disorder to an inverted configuration where the highest energy three-particle state corresponds to the right half qubits ($Q_4, Q_5, Q_6$). The conditional many-body evolution maps the solid + fluid state of the disorder-free system to a highly-entangled N00N state $\frac{1}{\sqrt{2}}\left(|L\rangle + |R\rangle\right)$ of the disordered system. In this shorthand notation, the N00N (Schr\"odinger cat) state corresponds to a superposition of having photons on the left half of the lattice (with an empty ancilla) $|L\rangle = |111\rangle |0\rangle |000\rangle$ and having photons on the right half of the lattice (with an excited ancilla) $|R\rangle = |000\rangle |1\rangle |111\rangle$.

The coherence of this long-range correlated state is characterized through the many-body Ramsey interferometry protocol~\cite{MBRamPaper} illustrated in Fig.~\ref{fig:adbNOON}b. Following the preparation of the entangled cat state, the system coherently evolves for a hold time $\Delta t$ to allow the left and right localized states to accumulate a relative phase $\frac{1}{\sqrt{2}}\left(|L\rangle + e^{i\Delta \phi}|R\rangle\right)$ proportional to the energy difference between the two entangled configurations $\Delta \phi = (\omega_R - \omega_L) \Delta t$. Finally, the reversed transport sequence is applied to disentangle the qubits and relocalize the information back into the ancilla qubit $|111\rangle \otimes \frac{1}{\sqrt{2}}\left(|0\rangle + e^{i\Delta \phi}|1\rangle \right) \otimes |000\rangle$. The accrued phase is now entirely encoded in the ancilla qubit, which we can extract through a subsequent $\frac{\pi}{2}$ pulse followed by a single-qubit measurement in the occupation basis $|0\rangle, |1\rangle$. 
Plotting the ancilla occupation versus the hold time yields the sinusoidal Ramsey fringes shown in Fig.~\ref{fig:adbNOON}c. These measured oscillations have a contrast limited by qubit dephasing (see SI~\ref{SI:SystemparametersandOperatingpoints}), and a frequency that is down-converted from the ancilla's natural precession frequency through a $\Delta t$-dependent phase offset on the second $\frac{\pi}{2}$ pulse (see SI~\ref{SI:RamseyInterferometryMeasurements}). The fringe frequency becomes the defining feature for validating entanglement: the frequency difference in the measured fringes when the ancilla qubit is uncoupled (no photons injected in the lattice) and entangled (transport protocol) with the lattice is equal to the energy difference between the left and right qubit clusters in the disordered configuration. The validation is twofold: the oscillations demonstrate a coherent superposition of multi-qubit eigenstates and the measured fringe frequency is consistent with the preparation of an entangled N00N state.
\newline
\indent We further validate the structure of multipartite entanglement in our synthetic material by benchmarking it as a \textit{sensor}, in this case for measuring an energy imbalance between sites to the left and right of our quantum-controlled switch. 
We demonstrate this principle by introducing external perturbations to the lattice during the Ramsey evolution $\Delta t$: the sites on the right (left) side of the ancilla are frequency shifted by a small offset $+(-) \delta$.
For uncorrelated lattice site occupations, this energy shift can be directly detected from individual Ramsey measurements on each qubit picking up the relative phase $\Delta\phi = \delta \Delta t$ in the measured fringe frequency.
Leveraging our entanglement preparation technique combined with the single-qubit Ramsey probe, we achieve an improved sensitivity from the collective enhancement of the acquired phase $\Delta\phi = (N-1)\delta \Delta t$ leading to an increased fringe frequency. The $N-1$ prefactor appears instead of $N$ because the ancilla frequency is deliberately kept constant during the perturbed Ramsey sequence. This measurement is conducted for $N=5$ and $N=7$ qubits, as shown in Fig.~\ref{fig:adbNOON}d. The results clearly highlight the collective shift $(N-1)\delta$ of the many-body Ramsey fringe frequency in response to the applied lattice perturbation, serving as a definitive signature of assembling the entangled N00N state. 

The entanglement generation protocol relies on the ability to adiabatically assemble and disassemble the solid + fluid many-body superposition. Similar to the adiabatic preparation of correlated fluids~\cite{Paredes2004TonksGirardeauLattice, AdbPaper, leonard2023realization}, this becomes a balancing act between evolving the system at a slow enough rate (relative to the energy gaps) to satisfy adiabaticity, and a fast enough rate to avoid particle loss/dephasing (set by the photon $T_1$/$T_2$ lifetimes). 
We identify this optimum ramp rate by monitoring the Fourier spectrum of the ancilla Ramsey fringes as a function of the ramp duration $t_\mathrm{ramp}$. We perform this measurement for an $N = 5$ site lattice as shown in Fig.~\ref{fig:adbNOON}e, where the ramp time is varied over two orders of magnitude. For very fast ramps $t_\mathrm{ramp} \ll J^{-1}$, the photons remain localized as they lack time to move at all, even when the transport is otherwise unimpeded. In this scenario, the ancilla is uncoupled from the other sites and is precessing at its reference frequency $\omega_\mathrm{ref}/2\pi\simeq 50\,\textrm{MHz}$ which coincides with the dominant frequency component in the measured Ramsey fringe. For intermediate ramps $t_\mathrm{ramp} \approx J^{-1}$, the photons do not adiabatically follow the conditioned trajectories in Fig.~\ref{fig:adbNOON}a, but diabatically delocalize to other fluid eigenstates.
These diabatic excitations in the ordered lattice are mapped to W-like entangled states in the disordered configuration, appearing as discrete frequency components in the Ramsey spectrum. Further increasing the ramp time $t_\mathrm{ramp} \gg J^{-1}$ ensures the photons adiabatically delocalize into the target fluid state and conditionally prepares the solid + fluid many-body state. This progression from a product state to the final entangled state is captured as a single frequency component prevailing in the measured Ramsey spectrum, corresponding to the fringe frequency $(\omega_\mathrm{ref} + \omega_R - \omega_L)/2\pi \simeq -250\,\textrm{MHz}$ for the N00N state. 
The measured signal is folded about the $500\,\textrm{MHz}$ Nyquist limit and thus appears as a frequency component at $250\,\textrm{MHz}$.
Evolving for longer ramp times, decoherence effects manifest as a decreased Ramsey fringe contrast and thus a diminished peak amplitude of the Fourier component associated with the entangled state.

\begin{figure}[htp] 
	\centering
	\includegraphics[width=0.5\textwidth]{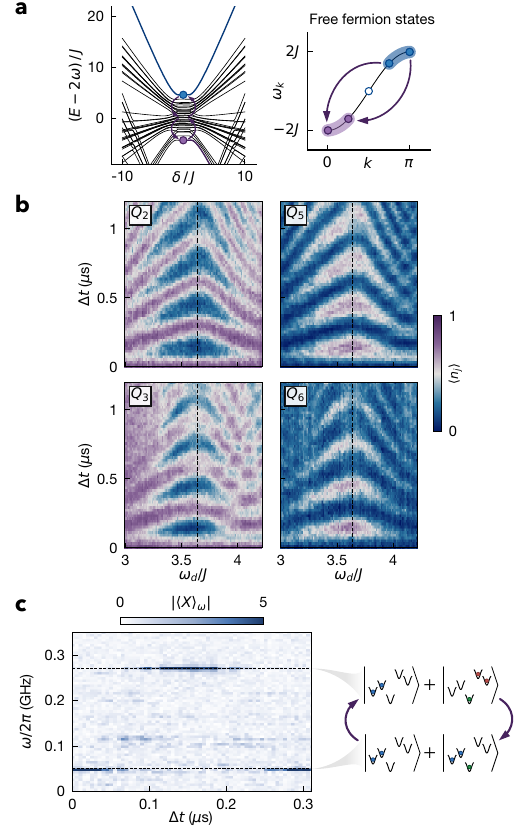}
	\caption[width=\textwidth]{
		\textbf{Phonon-assisted SWAP operation}. \textbf{a.} Entanglement generation using conditional phonon dynamics is demonstrated for $N=5$ qubits. Phonons are excited in the adiabatically prepared fluid through the potential modulation of a single lattice site (flux driving the transmon), creating a particle-hole excitation. Driving the lattice site at $\omega_d \approx 3.6 J$ induces a two-phonon process that inverts the population of the free fermion eigenstates (diagram). \textbf{b.} Adiabatically reintroducing disorder turns the phonon creation process into a many-body SWAP operation in the computational basis, where the highest frequency qubits ($Q_2,\, Q_3$) simultaneously exchange photons with the lowest frequency qubits ($Q_5,\, Q_6$). In contrast, this lattice perturbation does not affect the solid (Mott insulator) eigenstate, since it is incompressible ($\omega_d \ll U$). Therefore, applying this phonon drive to the $|\text{solid}\rangle + |\text{fluid}\rangle$ phase leads to a similar swap operation conditioned on the state of the ancilla ($Q_4$). \textbf{c.} The preparation of the target entangled state is clearly captured in a many-body Rabi experiment: the ancilla Ramsey frequency is probed as a function of the phonon drive duration, and we observe a cyclic transition from the product state to the entangled cat state.
  }\label{fig:phononSWAP}
\end{figure}

\section{Conditional Phonon Transport}
\label{sec:condphonontransport}
A complementary approach for creating cat states is inspired by the observation that they can be constructed through a conditional multi-qubit SWAP operation: photons from the left half of the lattice are transferred over the right half only if the middle ancilla qubit is excited. We demonstrate this protocol for a $N=5$ site lattice in Fig.~\ref{fig:phononSWAP}.
The SWAP operation itself leverages the one-to-one correspondence between the product states in the disordered system and the fluid eigenstates in the disorder-free transistor configuration, where photons localized on the left ($Q_2, Q_3$) and right ($Q_5, Q_6$) sides of the disordered lattice are adiabatically connected to the highest and lowest energy fluid eigenstates in the two-particle hard-core band, respectively.
The intuition gained from mapping fluid states of hardcore bosons to free fermions~\cite{cazalilla2011rmp}, as shown in Fig.~\ref{fig:phononSWAP}a, reveals that such few-body SWAP operations are equivalent to inverting the population of occupied \& empty free-fermion eigenstates. 
This population transfer is enabled by creating particle-hole excitations -- specifically \textit{phonons} -- corresponding to collective modes in the fluid.
These phonon excitations are generated by perturbing the lattice potential, by modulating the energy of a single lattice site  $\varepsilon_d \cos(\omega_d t) n_i$ (through flux driving the transmon) with a drive frequency $\omega_d$ (see SI~\ref{SI:PhononTransport}). In practice, we modulate the site $Q_3$ site at $\omega_d \approx 3.6 J$ (with a constant drive amplitude $\varepsilon_d$) to drive a two-phonon transition between the fluid states at the edges of the hard-core band. Adiabatically tuning the qubits back to their initial disordered configuration converts the phonon creation in the fluid basis to a multi-qubit SWAP operation in the computational basis. This process is highlighted in Fig.~\ref{fig:phononSWAP}b, where the higher frequency sites ($Q_2$, $Q_3$) are simultaneously exchanging photons with the lower frequency sites ($Q_5$, $Q_6$).

Implementing the ancilla-conditioned SWAP operation follows naturally from applying the phonon drive to our solid + fluid state prepared in the transistor configuration. In contrast to the fluid, the solid (Mott insulating) state is not affected by the lattice modulation since it is an \textit{incompressible} state with an interaction-induced gap larger than the phonon drive frequency ($U \gg \omega_d$). Benchmarking the operation is then simply a matter of  monitoring the $Q_4$ ancilla Ramsey fringe frequency as we vary the duration of the phonon drive, as shown in Fig.~\ref{fig:phononSWAP}c. This presents the multi-particle version of a Rabi oscillation, whereby we periodically cycle between the product $|11\rangle \frac{1}{\sqrt{2}}(|0\rangle + |1\rangle) |00\rangle$ and entangled $\frac{1}{\sqrt{2}}(|11\rangle|0\rangle|00\rangle + |00\rangle|1\rangle|11\rangle)$ states.
This technique of optimizing single-qubit drive parameters (for inducing phonons) with respect to a single qubit observable (ancilla fringes) highlights the versatility of our platform for implementing few-body entangling operations.

\begin{figure*}[ht] 
	\centering
	\includegraphics[width=1\textwidth]{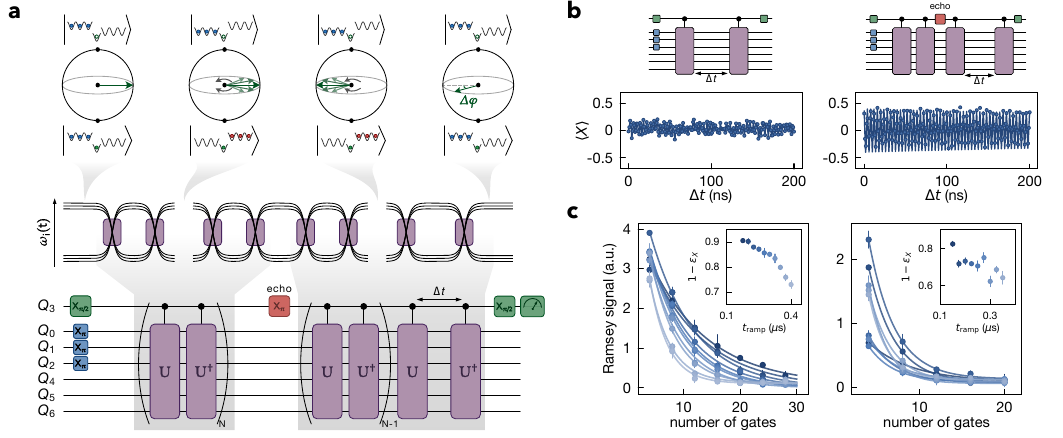}
	\caption{
	\textbf{Many-body echo}. \textbf{a.} An echo protocol is employed to dynamically decouple low-frequency (compared to the evolution time) phase noise during the adiabatic evolution of the entangling operation. The control sequence is an extension of the many-body Ramsey procedure, applying pairs of entangling ($U_\textrm{MB}$) and disentangling ($U_\textrm{MB}^\dagger$) operations by adiabatically modulating the qubit-site frequencies $\omega_i(t)$. The many-body phase information $\Delta \phi$ is relocalized in the ancilla qubit by inserting the Ramsey hold time $\Delta t$ prior to the last disentangling operation. The low-frequency phase noise, accrued during the adiabatic ramps, is averaged out by driving the ancilla with a \textpi-pulse in the middle of the sequence. \textbf{b.} Dynamically refocusing the ancilla phase information enhances the measured Ramsey fringes, as demonstrated for the seven-qubit N00N state: a standard Ramsey sequence (left) yields no detectable coherence, whereas an echo sequence (right) yields visible fringe oscillations. \textbf{c.} We quantify the fidelity of preparing the off-diagonal elements of the density matrix as a direct measure of coherence. The extended echo sequence is used for amplifying and estimating the off-diagonal error $\epsilon_X$ in each entangling/disentangling operation by measuring the decay in the ancilla echo signal (magnitude of the Fourier component) $\langle X \rangle_\omega=A(1-\epsilon_X)^{4N}$. For the five- and seven-qubit N00N states we measure an average error of $\epsilon_X =(9.1\pm0.9)\%$ and $(17.7\pm 1.3)\%$, respectively. Preparation errors further increase as we increase the duration of the adiabatic ramp (inset plots).
 }
    \label{fig:MBecho}
\end{figure*}

\section{Many-body echo}
\label{sec:mbecho}

While N00N states exhibit greater resilience to correlated phase noise compared to Greenberger-Horne-Zeilinger (GHZ) states~\cite{Monz_PRL2011}, they still remain vulnerable to losses and dephasing as the system size grows, causing the scaling of the phase sensitivity to degrade from the Heisenberg limit to the standard quantum limit~\cite{Escher_NatPhys2011, Bohmann_PRA2015}. 
In our protocol, the entanglement-enhanced phase accumulation (set by the energy imbalance between the two cat components) is an order of magnitude faster than the photon transport (set by the tunneling rate $J$).
Consequently, the system spends more time adiabatically traversing the many-body eigenstate trajectories (see Fig.~\ref{fig:adbNOON}a) than performing entanglement-enhanced sensing in the N00N state.
The fidelity of preparing and probing these entangled states is thus constrained by decoherence during the adiabatic ramps, which, in the case of our flux-tunable transmon sites, is limited by dephasing noise with $T_2\approx 4$\textmu s.

The hybrid nature of our platform -- which allows for coupling quantum computers to many-body systems -- provides a unique opportunity to exploit control tools to enhance the coherence of many-body states.
Specifically, we circumvent the decoherence bottleneck tied to the preparation of our N00N states by employing an echo protocol tailored to dynamically decouple the many-body evolution from low-frequency phase noise, thereby refocusing the ancilla phase information~\cite{Bylander_NatPhys2011}.
Echo protocols are also actively investigated for preparing many-body entangled states in Ising spin chains and Rydberg atom arrays~\cite{zeng2025adiabatic}.

Our many-body echo protocol, displayed in Fig.~\ref{fig:MBecho}a, represents a generalization of the Ramsey sequence (Fig.~\ref{fig:adbNOON}b). 
The system evolves in a series of alternating entangling ($U_\textrm{MB}$) and disentangling ($U_\textrm{MB}^\dagger$) transport sequences, inserting a hold time $\Delta t$ in between the final pair of operations to accumulate the many-body phase information in the ancilla qubit.
The key feature of this echo procedure is the application of a \textpi-pulse on the ancilla qubit in the middle of the sequence, akin to a Hahn spin echo~\cite{Hahn_PR1950}.
This inversion refocuses the delicate phase information in the ancilla qubit by effectively reversing the accumulated phase error arising from shot-to-shot flux noise during the adiabatic ramps.
Applying a single pair of entangling and disentangling operations before and after the ancilla \textpi-rotation is sufficient to dynamically refocus the ancilla phase information and thus enhance its coherence, as demonstrated by the measured Ramsey fringes for the five- and seven-qubit N00N states.
While the echo pulse results in noticeably higher contrast oscillations for the five-qubit state, its effect is most pronounced in the seven-qubit state, shown in Fig.~\ref{fig:MBecho}b, where coherence is undetectable with the standard Ramsey sequence and fringe oscillations emerge only when the echo is applied.

To characterize the performance of the entangling protocol, we quantify the coherence of the prepared N00N state associated with the off-diagonal elements of its density matrix. The coherence is accessible from the amplitude of the measured many-body Ramsey fringes, analogous to probing parity oscillations in GHZ states~\cite{Monz_PRL2011}.
Using error amplification, we apply repeated pairs of entangling and disentangling operations to estimate the off-diagonal error of a single entangling operation.
The extended echo sequence displayed in Fig.~\ref{fig:MBecho}a is used for estimating the average off-diagonal error $\epsilon_X$ by fitting the decay in the measured ancilla echo signal to $\langle X \rangle_\omega (N) = A(1-\epsilon_X)^{4N}$, where $N$ is the number of entangling-disentangling pair operations applied prior to the echo pulse. As shown in Fig.~\ref{fig:MBecho}c, we measure an average error rate of $\epsilon_X =(9.1\pm0.9)\%$ for the five-qubit N00N state, and $\epsilon_X =(17.7\pm 1.3)\%$ for the seven-qubit state. 
Additionally, we estimate the errors associated with the non-adiabatic evolution of our many-body system by probing the fidelity of reversing the entanglement generation. The sequence, detailed in SI~\ref{SI:PulseSequences}, involves applying multiple pairs of entangling and disentangling operations and measuring the probability of returning the system to its initial product state by probing the final occupation of the lattice sites initially injected with photons. The average reversibility error $\epsilon_\mathrm{rev}$ is obtained from fitting the fidelity metric to $\mathcal{F}_\mathrm{adb}(N) = A(1-\epsilon_\mathrm{rev})^{2N}$. We extract error rates of $(5.4 \pm 0.2)\%$ and $(4.7 \pm 0.2)\%$ for the five- and seven-qubit entangled states. These errors closely match the values limited by single-qubit decoherence, extracted from decoupling the entangling dynamics by running the same sequence with the ancilla qubit in the ground state.

\section{Conclusion and Outlook}
\label{sec:outlook}

In this work we have explored the consequences of using a minimal quantum computer to control the evolution of a quantum many-body systems To realize this control, we have devised a new protocol that employs ancilla-conditioned dynamics to steer these synthetic many-body systems to unconventional strongly-correlated states where different phases of matter coexist in superposition.
Instead of adopting the conventional approach of entangling an ancilla qubit to the initial state~\cite{googlescramble2021, GoogleAnyons2021, roberts2024manybody}, we entangle the ancilla to the Hamiltonian degrees of freedom that drive the coherent evolution, as this approach can navigate larger sectors of the Hilbert space. This Hamiltonian-level entanglement with an ancilla qubit further facilitates the efficient extraction of information reflecting the energies of the constituent many-body states, from a Ramsey measurement of the accumulated phase localized in the ancilla qubit.

Our protocol entangles the potential profile of a 1D tight-binding lattice of capacitively coupled transmon qubits, implementing a quantum-controlled transistor where a tunnel-barrier regulating transport of microwave photons along the lattice depends on the state of an ancilla qubit. Injecting photons on one side of the lattice in a Mott insulating state and setting the ancilla qubit in $\frac{1}{\sqrt{2}}(|0\rangle + |1\rangle )$ allows the entire many-body system to evolve to a $\frac{1}{\sqrt{2}}(|\text{solid}\rangle + |\text{fluid}\rangle )$ state which is subsequently adiabatically mapped to a highly-entangled cat state in the computational basis of the transmon lattice sites. Combining this ancilla-conditioned entangling operation with its time-reversed counterpart into a Ramsey sequence enables the direct measurement of long-range coherence from the ancilla Ramsey fringe. The fringe frequency is proportional to the energy difference between the multi-qubit cat components, acting as a witness for detecting and quantifying the multipartite entanglement in our many-body system using concepts typically employed in metrology~\cite{hauke2016measuring}. 
\newline
\indent Marrying ancilla-conditioned many-body operations with Ramsey spectroscopy provides a powerful framework for various key applications in quantum science. With a larger control register, such an approach could enhance the sensitivity and dynamic range of many-body sensors by embedding quantum Fourier transforms into the entangling dynamics~\cite{vorobyov2021quantum}. Generalizations of our echo-based dynamical decoupling scheme, combined with multi-qubit operations, suggest a new  quantum signal processing algorithms~\cite{motlagh2024generalized}. Furthermore, this protocol can be adapted to measure the entanglement spectrum of many-body states from Ramsey spectroscopy of a single control qubit that conditions the global swap between two copies of the state~\cite{pichler2016measurement}. 
More broadly, this work demonstrates that integrating analog many-body systems with small quantum computers creates a powerful synergy, offering new avenues for efficiently probing properties of synthetic matter and generating large-scale entanglement in information processors.

\section{Acknowledgments}
This work was supported by ARO MURI Grant W911NF-15-1-0397, AFOSR MURI Grant FA9550-19-1-0399, and by NSF Eager Grant 1926604. Support was also provided by the Chicago MRSEC, which is funded by NSF through Grant DMR-1420709. A.V. acknowledges support from the MRSEC-funded Kadanoff-Rice Postdoctoral Research Fellowship. G.R. acknowledges support from the NSF GRFP. 
K.R.A.H. acknowledges support from the National Science
Foundation (PHY-1848304, DGE-2346014), the Department of Energy (DE-SC0024301), the W. M. Keck Foundation
(Grant No. 995764), and the Office of Naval Research (N00014-20-1-2695).
We acknowledge support from the Samsung Advanced Institute of Technology Global Research Partnership. Devices were fabricated in the Pritzker Nanofabrication Facility at the University of Chicago, which receives support from Soft and Hybrid Nanotechnology Experimental (SHyNE) Resource (NSF ECCS-1542205), a node of the National Science Foundation’s National Nanotechnology Coordinated Infrastructure.
\newline
\section{Author Contributions}
The experiments were designed by A.V., G.R., J.S., and D.I.S. The collection of data was handled by A.V. Theoretical support was provided by K.H. All authors analyzed the data and contributed to the manuscript.








%
\section{Methods}

\textbf{ Device Fabrication }

\noindent The superconducting circuit was fabricated on a 10$\times$20 mm C-plane sapphire chip. Prior to deposition, the $450~\mu\text{m}$ thick sapphire wafer was subjected to a thorough cleaning procedure, which included annealing at $1500^{\circ}\text{C}$ for 2 hours, sequential solvent cleaning, and etching in heated $80^{\circ}\text{C}$ Nano-Strip and $40^{\circ}\text{C}$ sulfuric acid to ensure a pristine surface.
The device features were patterned in two lithography steps. First, optical lithography was employed to define the large-scale components, including the wiring and ground plane, in a $200~\text{nm}$ thick tantalum (Ta) base layer. This Ta film was deposited at $800^{\circ}\text{C}$ and then patterned using a direct pattern writer (Heidelberg MLA 150) followed by wet etching with hydrofluoric acid (HF).The second step involved electron-beam (e-beam) lithography to define the critical, small-scale features: the aluminum (Al) Josephson junctions (JJs) and superconducting quantum interference device (SQUID) loops. A MMA-PMMA bilayer resist was exposed using a Raith EBPG5000 Plus E-Beam Writer.
The $\text{Al}/\text{AlO}_{\text{x}}/\text{Al}$ Manhattan-style JJs were formed via a shadow evaporation technique in an angled e-beam evaporator (Plassys MEB550). Before the initial Al deposition, Argon (Ar) ion milling was performed on the exposed Ta features to remove any native Ta oxide, guaranteeing an improved ohmic contact between the Ta base layer and the superconducting Al layers. The first $60~\text{nm}$ Al layer was evaporated at $0.1~\text{nm/s}$ at an angle of $30^{\circ}$ relative to the substrate normal. The oxidation barrier was formed by exposing the first Al layer to $\text{O}_2$ gas for 24 minutes at a pressure of $50~\text{mBar}$. The final $150~\text{nm}$ Al layer was evaporated at the same rate and $30^{\circ}$ angle, but with the substrate rotated $90^{\circ}$ in-plane relative to the first evaporation, completing the overlapped junction structure.

\textbf{ Device Parameters }

\noindent The Bose-Hubbard Hamiltonian for microwave photons is realized in a one-dimensional chain of capacitively coupled transmon qubits. The device is engineered to achieve a nearest-neighbor photon tunneling energy of $J/2\pi \approx 9~\text{MHz}$ via capacitive coupling. The intrinsic transmon nonlinearity is used to provide a strong photon-photon interaction energy of $U/2\pi \approx 240~\text{MHz}$. These parameters were optimized through finite element simulations (Ansys HFSS) and experimental iteration. Each qubit is connected to a dedicated readout resonator and Purcell filter for state detection. A table of system parameters and further device details are available in section SI~\ref{SI:SystemparametersandOperatingpoints}.
The optimal qubit lattice frequency was periodically adjusted between $4.8-5.3~\text{GHz}$ throughout the experiment to mitigate the influence of two-level system (TLSs) defects.
Coherence times exhibited temporal variation, with typical averages of $T_1 \approx 45~\mu s$ and $T_2^{*} \approx 1.5~\mu s$. Operating all qubits at resonance provides an effective flux sweet-spot due to avoided crossings, which is expected to enhance $T_2$ during the disorder-free lattice evolution.

\textbf{ Microwave Wiring }

The superconducting chip is mounted and heavily wire-bonded to a multilayer copper PCB. Dense wire-bonding around the device ensures a continuous ground plane and suppresses spurious slotline modes. The PCB assembly is secured within an oxygen-free high conductivity (OFHC) copper mount. This packaged sample is then mounted to the mixing chamber plate of a dilution refrigerator operating at a base temperature of approximately $8-9~\text{mK}$. To prevent external noise and radiation from interfering with the device, the sample is enclosed in extensive shielding, typically comprising nested layers of copper, superconducting lead, and MuMetal. Control microwave signals are routed from the room-temperature measurement setup through microwave coaxial cables. DC signals used for flux biasing the transmon SQUIDs are routed through twisted pair lines connected to individual on-chip lines inductively coupled to each SQUID. A separate solenoid of coiled niobium-titanium (NbTi) wire, mounted to the sample package, provides a global magnetic field used for tuning the qubits close to the desired frequency configuration, with minimal heating compared to driving the individual flux bias lines. Superconducting NbTi coaxial lines carry the multiplexed output signal of the readout resonators from the device feedline to the room-temperature homodyne measurement setup, going through a low-noise high-electron-mobility transistor (HEMT) amplifier mounted to the $4~\text{K}$ plate of the dilution refrigerator. See the supplemental information of our previous work~\cite{AdbPaper} for further details on the cryogenic and room-temperature wiring.

\textbf{ Flux Control and Crosstalk }

We control in real-time the frequencies of the transmon qubits by biasing their respective SQUID loops with magnetic fields supplied with currents sent through the inductively coupled bias lines. The frequency range $\omega_{01}/2\pi \approx (4-6)\,\text{GHz}$ allows us to tune the lattice potential between different ordered and disordered configuration (see SI~\ref{SI:SystemparametersandOperatingpoints}). The mutual inductive coupling with other SQUID loops creates non-negligible cross-talk, which we address by measuring and inverting a calibrated cross-talk matrix for DC and RF flux signals. Additional details are available in section SI~\ref{SI:FluxControlandCrosstalk}

\textbf{ Readout }

On-site microscopy is done by sending microwave tones through the device feedline to probe the response of individual readout resonators dispersively coupled to each transmon qubit. Assuming the qubit populations are restricted in the $(|0\rangle, |1\rangle, |2\rangle)$ manifold, we probe the readout resonators at two separate frequencies to maximally distinguish the states $|0\rangle$ from $|1\rangle, |2\rangle$, and $|1\rangle$ from $|0\rangle, |2\rangle$. Our single-shot readout fidelities are in the range $85-95\%$. The non-negligible binning errors are corrected with a calibrated confusion matrix.





\subsection{Data Availability}
The experimental data presented in this manuscript are available from the corresponding author upon request, due to the proprietary file formats employed in the data collection process.

\subsection{Code Availability}
The source code for simulations throughout are available from the corresponding author upon request. 

\subsection{Additional Information}
Correspondence and requests for materials should be addressed to A.V. (andrei.v@nyu.edu) and D.S. (dschus@stanford.edu). Supplementary information is available for this paper.

\newpage
\clearpage

\bibliographystyle{naturemag}
\bibliography{references}

@article{Monz_PRL2011,
  title = {14-Qubit Entanglement: Creation and Coherence},
  author = {Monz, Thomas and Schindler, Philipp and Barreiro, Julio T. and Chwalla, Michael and Nigg, Daniel and Coish, William A. and Harlander, Maximilian and H\"ansel, Wolfgang and Hennrich, Markus and Blatt, Rainer},
  journal = {Phys. Rev. Lett.},
  volume = {106},
  issue = {13},
  pages = {130506},
  numpages = {4},
  year = {2011},
  month = {Mar},
  publisher = {American Physical Society},
  doi = {10.1103/PhysRevLett.106.130506}
}

@Article{Escher_NatPhys2011,
author={Escher, B. M.
and de Matos Filho, R. L.
and Davidovich, L.},
title={General framework for estimating the ultimate precision limit in noisy quantum-enhanced metrology},
journal={Nature Physics},
year={2011},
month={May},
day={01},
volume={7},
number={5},
pages={406-411},
issn={1745-2481},
doi={10.1038/nphys1958}
}

@article{Bohmann_PRA2015,
  title = {Entanglement and phase properties of noisy NOON states},
  author = {Bohmann, M. and Sperling, J. and Vogel, W.},
  journal = {Phys. Rev. A},
  volume = {91},
  issue = {4},
  pages = {042332},
  numpages = {8},
  year = {2015},
  month = {Apr},
  publisher = {American Physical Society},
  doi = {10.1103/PhysRevA.91.042332}
}

@Article{Bylander_NatPhys2011,
author={Bylander, Jonas
and Gustavsson, Simon
and Yan, Fei
and Yoshihara, Fumiki
and Harrabi, Khalil
and Fitch, George
and Cory, David G.
and Nakamura, Yasunobu
and Tsai, Jaw-Shen
and Oliver, William D.},
title={Noise spectroscopy through dynamical decoupling with a superconducting flux qubit},
journal={Nature Physics},
year={2011},
month={Jul},
day={01},
volume={7},
number={7},
pages={565-570},
issn={1745-2481},
doi={10.1038/nphys1994}
}

@article{Hahn_PR1950,
  title = {Spin Echoes},
  author = {Hahn, E. L.},
  journal = {Phys. Rev.},
  volume = {80},
  issue = {4},
  pages = {580--594},
  numpages = {0},
  year = {1950},
  month = {Nov},
  publisher = {American Physical Society},
  doi = {10.1103/PhysRev.80.580}
}

@article{MBRamPaper,
  title={Manybody interferometry of quantum fluids},
  author={Roberts, Gabrielle and Vrajitoarea, Andrei and Saxberg, Brendan and Panetta, Margaret G and Simon, Jonathan and Schuster, David I},
  journal={Science Advances},
  volume={10},
  number={29},
  pages={eado1069},
  year={2024},
  publisher={American Association for the Advancement of Science}
}

@article{vorobyov2021quantum,
  title={Quantum Fourier transform for nanoscale quantum sensing},
  author={Vorobyov, Vadim and Zaiser, Sebastian and Abt, Nikolas and Meinel, Jonas and Dasari, Durga and Neumann, Philipp and Wrachtrup, J{\"o}rg},
  journal={npj Quantum Information},
  volume={7},
  number={1},
  pages={124},
  year={2021},
  publisher={Nature Publishing Group UK London}
}

@article{landsman2019verified,
  title={Verified quantum information scrambling},
  author={Landsman, Kevin A and Figgatt, Caroline and Schuster, Thomas and Linke, Norbert M and Yoshida, Beni and Yao, Norman Y and Monroe, Christopher},
  journal={Nature},
  volume={567},
  number={7746},
  pages={61--65},
  year={2019},
  publisher={Nature Publishing Group UK London}
}

@article{leonard2023realization,
  title={Realization of a fractional quantum Hall state with ultracold atoms},
  author={L{\'e}onard, Julian and Kim, Sooshin and Kwan, Joyce and Segura, Perrin and Grusdt, Fabian and Repellin, C{\'e}cile and Goldman, Nathan and Greiner, Markus},
  journal={Nature},
  pages={1--5},
  year={2023},
  publisher={Nature Publishing Group UK London}
}

@article{kollar2019hyperbolic,
  title={Hyperbolic lattices in circuit quantum electrodynamics},
  author={Koll{\'a}r, Alicia J and Fitzpatrick, Mattias and Houck, Andrew A},
  journal={Nature},
  volume={571},
  number={7763},
  pages={45--50},
  year={2019},
  publisher={Nature Publishing Group UK London}
}

@article{AdbPaper,
	title = {Disorder-assisted assembly of strongly correlated fluids of light},
	volume = {616},
	doi = {10.1038/s41586-022-05357-x},
	journal = {Nature},
	author = {Saxberg, Brendan and Vrajitoarea, Andrei and Roberts, Gabrielle and Panetta, Margaret and Simon, Jonathan and Schuster, David I},
	month = dec,
	year = {2022},
	pages = {435-441},
}

@article{Morvan2022-boundstate,
  title    = "Formation of robust bound states of interacting microwave photons",
  author   = "Morvan, A and Andersen, T I and Mi, X and Neill, C and Petukhov,
              A and Kechedzhi, K and Abanin, D A and Michailidis, A and
              Acharya, R and Arute, F and Arya, K and Asfaw, A and Atalaya, J
              and Bardin, J C and Basso, J and Bengtsson, A and Bortoli, G and
              Bourassa, A and Bovaird, J and Brill, L and Broughton, M and
              Buckley, B B and Buell, D A and Burger, T and Burkett, B and
              Bushnell, N and Chen, Z and Chiaro, B and Collins, R and Conner,
              P and Courtney, W and Crook, A L and Curtin, B and Debroy, D M
              and Del Toro Barba, A and Demura, S and Dunsworth, A and Eppens,
              D and Erickson, C and Faoro, L and Farhi, E and Fatemi, R and
              Flores Burgos, L and Forati, E and Fowler, A G and Foxen, B and
              Giang, W and Gidney, C and Gilboa, D and Giustina, M and Grajales
              Dau, A and Gross, J A and Habegger, S and Hamilton, M C and
              Harrigan, M P and Harrington, S D and Hoffmann, M and Hong, S and
              Huang, T and Huff, A and Huggins, W J and Isakov, S V and
              Iveland, J and Jeffrey, E and Jiang, Z and Jones, C and Juhas, P
              and Kafri, D and Khattar, T and Khezri, M and Kieferov{\'a}, M
              and Kim, S and Kitaev, A Y and Klimov, P V and Klots, A R and
              Korotkov, A N and Kostritsa, F and Kreikebaum, J M and Landhuis,
              D and Laptev, P and Lau, K-M and Laws, L and Lee, J and Lee, K W
              and Lester, B J and Lill, A T and Liu, W and Locharla, A and
              Malone, F and Martin, O and McClean, J R and McEwen, M and Meurer
              Costa, B and Miao, K C and Mohseni, M and Montazeri, S and Mount,
              E and Mruczkiewicz, W and Naaman, O and Neeley, M and Nersisyan,
              A and Newman, M and Nguyen, A and Nguyen, M and Niu, M Y and
              O'Brien, T E and Olenewa, R and Opremcak, A and Potter, R and
              Quintana, C and Rubin, N C and Saei, N and Sank, D and
              Sankaragomathi, K and Satzinger, K J and Schurkus, H F and
              Schuster, C and Shearn, M J and Shorter, A and Shvarts, V and
              Skruzny, J and Smith, W C and Strain, D and Sterling, G and Su, Y
              and Szalay, M and Torres, A and Vidal, G and Villalonga, B and
              Vollgraff-Heidweiller, C and White, T and Xing, C and Yao, Z and
              Yeh, P and Yoo, J and Zalcman, A and Zhang, Y and Zhu, N and
              Neven, H and Bacon, D and Hilton, J and Lucero, E and Babbush, R
              and Boixo, S and Megrant, A and Kelly, J and Chen, Y and
              Smelyanskiy, V and Aleiner, I and Ioffe, L B and Roushan, P",
  journal  = "Nature",
  volume   =  612,
  number   =  7939,
  pages    = "240--245",
  month    =  dec,
  year     =  2022
}

@article{GoogleAnyons2021,
author = {K. J. Satzinger  and Y.-J Liu  and A. Smith  and C. Knapp  and M. Newman  and C. Jones  and Z. Chen  and C. Quintana  and X. Mi  and A. Dunsworth  and C. Gidney  and I. Aleiner  and F. Arute  and K. Arya  and J. Atalaya  and R. Babbush  and J. C. Bardin  and R. Barends  and J. Basso  and A. Bengtsson  and A. Bilmes  and M. Broughton  and B. B. Buckley  and D. A. Buell  and B. Burkett  and N. Bushnell  and B. Chiaro  and R. Collins  and W. Courtney  and S. Demura  and A. R. Derk  and D. Eppens  and C. Erickson  and L. Faoro  and E. Farhi  and A. G. Fowler  and B. Foxen  and M. Giustina  and A. Greene  and J. A. Gross  and M. P. Harrigan  and S. D. Harrington  and J. Hilton  and S. Hong  and T. Huang  and W. J. Huggins  and L. B. Ioffe  and S. V. Isakov  and E. Jeffrey  and Z. Jiang  and D. Kafri  and K. Kechedzhi  and T. Khattar  and S. Kim  and P. V. Klimov  and A. N. Korotkov  and F. Kostritsa  and D. Landhuis  and P. Laptev  and A. Locharla  and E. Lucero  and O. Martin  and J. R. McClean  and M. McEwen  and K. C. Miao  and M. Mohseni  and S. Montazeri  and W. Mruczkiewicz  and J. Mutus  and O. Naaman  and M. Neeley  and C. Neill  and M. Y. Niu  and T. E. O’Brien  and A. Opremcak  and B. Pató  and A. Petukhov  and N. C. Rubin  and D. Sank  and V. Shvarts  and D. Strain  and M. Szalay  and B. Villalonga  and T. C. White  and Z. Yao  and P. Yeh  and J. Yoo  and A. Zalcman  and H. Neven  and S. Boixo  and A. Megrant  and Y. Chen  and J. Kelly  and V. Smelyanskiy  and A. Kitaev  and M. Knap  and F. Pollmann  and P. Roushan },
title = {Realizing topologically ordered states on a quantum processor},
journal = {Science},
volume = {374},
number = {6572},
pages = {1237-1241},
year = {2021},
doi = {10.1126/science.abi8378}}

@article{Koch2007,
  title = {Charge-insensitive qubit design derived from the Cooper pair box},
  author = {Koch, Jens and Yu, Terri M. and Gambetta, Jay and Houck, A. A. and Schuster, D. I. and Majer, J. and Blais, Alexandre and Devoret, M. H. and Girvin, S. M. and Schoelkopf, R. J.},
  journal = {Phys. Rev. A},
  volume = {76},
  issue = {4},
  pages = {042319},
  numpages = {19},
  year = {2007},
  month = {Oct},
  publisher = {American Physical Society},
  doi = {10.1103/PhysRevA.76.042319}
}

@article{Cazalilla_Tonks_cnt_lat,
  title     = "Differences between the Tonks regimes in the continuum and on
               the lattice",
  author    = "Cazalilla, M A",
  journal   = "Phys. Rev. A",
  publisher = "American Physical Society",
  volume    =  70,
  number    =  4,
  pages     = "041604",
  month     =  oct,
  year      =  2004
}

@article{googlescramble2021,
	author = {Xiao Mi and Pedram Roushan and Chris Quintana and Salvatore Mandr{\`a} and Jeffrey Marshall and Charles Neill and Frank Arute and Kunal Arya and Juan Atalaya and Ryan Babbush and Joseph C. Bardin and Rami Barends and Joao Basso and Andreas Bengtsson and Sergio Boixo and Alexandre Bourassa and Michael Broughton and Bob B. Buckley and David A. Buell and Brian Burkett and Nicholas Bushnell and Zijun Chen and Benjamin Chiaro and Roberto Collins and William Courtney and Sean Demura and Alan R. Derk and Andrew Dunsworth and Daniel Eppens and Catherine Erickson and Edward Farhi and Austin G. Fowler and Brooks Foxen and Craig Gidney and Marissa Giustina and Jonathan A. Gross and Matthew P. Harrigan and Sean D. Harrington and Jeremy Hilton and Alan Ho and Sabrina Hong and Trent Huang and William J. Huggins and L. B. Ioffe and Sergei V. Isakov and Evan Jeffrey and Zhang Jiang and Cody Jones and Dvir Kafri and Julian Kelly and Seon Kim and Alexei Kitaev and Paul V. Klimov and Alexander N. Korotkov and Fedor Kostritsa and David Landhuis and Pavel Laptev and Erik Lucero and Orion Martin and Jarrod R. McClean and Trevor McCourt and Matt McEwen and Anthony Megrant and Kevin C. Miao and Masoud Mohseni and Shirin Montazeri and Wojciech Mruczkiewicz and Josh Mutus and Ofer Naaman and Matthew Neeley and Michael Newman and Murphy Yuezhen Niu and Thomas E. O'Brien and Alex Opremcak and Eric Ostby and Balint Pato and Andre Petukhov and Nicholas Redd and Nicholas C. Rubin and Daniel Sank and Kevin J. Satzinger and Vladimir Shvarts and Doug Strain and Marco Szalay and Matthew D. Trevithick and Benjamin Villalonga and Theodore White and Z. Jamie Yao and Ping Yeh and Adam Zalcman and Hartmut Neven and Igor Aleiner and Kostyantyn Kechedzhi and Vadim Smelyanskiy and Yu Chen},
	doi = {10.1126/science.abg5029},
	journal = {Science},
	number = {6574},
	pages = {1479-1483},
	title = {Information scrambling in quantum circuits},
	volume = {374},
	year = {2021}
}

@article{brown2019bad,
  title={Bad metallic transport in a cold atom Fermi-Hubbard system},
  author={Brown, Peter T and Mitra, Debayan and Guardado-Sanchez, Elmer and Nourafkan, Reza and Reymbaut, Alexis and H{\'e}bert, Charles-David and Bergeron, Simon and Tremblay, A-MS and Kokalj, Jure and Huse, David A and others},
  journal={Science},
  volume={363},
  number={6425},
  pages={379--382},
  year={2019},
  publisher={American Association for the Advancement of Science}
}

@article{eisert2015quantum,
  title={Quantum many-body systems out of equilibrium},
  author={Eisert, Jens and Friesdorf, Mathis and Gogolin, Christian},
  journal={Nature Physics},
  volume={11},
  number={2},
  pages={124--130},
  year={2015},
  publisher={Nature Publishing Group}
}

@article{simon2011quantum,
  title={Quantum simulation of antiferromagnetic spin chains in an optical lattice},
  author={Simon, Jonathan and Bakr, Waseem S and Ma, Ruichao and Tai, M Eric and Preiss, Philipp M and Greiner, Markus},
  journal={Nature},
  volume={472},
  number={7343},
  pages={307--312},
  year={2011},
  publisher={Nature Publishing Group}
}

@article{bluvstein2021controlling,
  title={Controlling quantum many-body dynamics in driven Rydberg atom arrays},
  author={Bluvstein, Dolev and Omran, Ahmed and Levine, Harry and Keesling, Alexander and Semeghini, Giulia and Ebadi, Sepehr and Wang, Tout T and Michailidis, Alexios A and Maskara, Nishad and Ho, Wen Wei and others},
  journal={Science},
  volume={371},
  number={6536},
  pages={1355--1359},
  year={2021},
  publisher={American Association for the Advancement of Science}
}

@article{bakr2009quantum,
  title={A quantum gas microscope for detecting single atoms in a Hubbard-regime optical lattice},
  author={Bakr, Waseem S and Gillen, Jonathon I and Peng, Amy and F{\"o}lling, Simon and Greiner, Markus},
  journal={Nature},
  volume={462},
  number={7269},
  pages={74--77},
  year={2009},
  publisher={Nature Publishing Group}
}

@article{carusotto2020photonic,
  title={Photonic materials in circuit quantum electrodynamics},
  author={Carusotto, Iacopo and Houck, Andrew A and Koll{\'a}r, Alicia J and Roushan, Pedram and Schuster, David I and Simon, Jonathan},
  journal={Nature Physics},
  volume={16},
  number={3},
  pages={268--279},
  year={2020},
  publisher={Nature Publishing Group}
}

@article{bloch2012quantum,
  title={Quantum simulations with ultracold quantum gases},
  author={Bloch, Immanuel and Dalibard, Jean and Nascimbene, Sylvain},
  journal={Nature Physics},
  volume={8},
  number={4},
  pages={267--276},
  year={2012},
  publisher={Nature Publishing Group}
}

@article{blatt2012quantum,
  title={Quantum simulations with trapped ions},
  author={Blatt, Rainer and Roos, Christian F},
  journal={Nature Physics},
  volume={8},
  number={4},
  pages={277--284},
  year={2012},
  publisher={Nature Publishing Group}
}

@article{monroe2021programmable,
  title={Programmable quantum simulations of spin systems with trapped ions},
  author={Monroe, Christopher and Campbell, Wes C and Duan, L-M and Gong, Z-X and Gorshkov, Alexey V and Hess, Paul W and Islam, Rajibul and Kim, Kihwan and Linke, Norbert M and Pagano, Guido and others},
  journal={Reviews of Modern Physics},
  volume={93},
  number={2},
  pages={025001},
  year={2021},
  publisher={APS}
}

@article{daley2022practical,
  title={Practical quantum advantage in quantum simulation},
  author={Daley, Andrew J and Bloch, Immanuel and Kokail, Christian and Flannigan, Stuart and Pearson, Natalie and Troyer, Matthias and Zoller, Peter},
  journal={Nature},
  volume={607},
  number={7920},
  pages={667--676},
  year={2022},
  publisher={Nature Publishing Group UK London}
}

@article{trivedi2024quantum,
  title={Quantum advantage and stability to errors in analogue quantum simulators},
  author={Trivedi, Rahul and Franco Rubio, Adrian and Cirac, J Ignacio},
  journal={Nature Communications},
  volume={15},
  number={1},
  pages={6507},
  year={2024},
  publisher={Nature Publishing Group UK London}
}

@article{roushan2017spectroscopic,
  title={Spectroscopic signatures of localization with interacting photons in superconducting qubits},
  author={Roushan, Pedram and Neill, Charles and Tangpanitanon, J and Bastidas, Victor M and Megrant, A and Barends, Rami and Chen, Yu and Chen, Z and Chiaro, B and Dunsworth, A and others},
  journal={Science},
  volume={358},
  number={6367},
  pages={1175--1179},
  year={2017},
  publisher={American Association for the Advancement of Science}
}

@article{choi2016exploring,
  title={Exploring the many-body localization transition in two dimensions},
  author={Choi, Jae-yoon and Hild, Sebastian and Zeiher, Johannes and Schau{\ss}, Peter and Rubio-Abadal, Antonio and Yefsah, Tarik and Khemani, Vedika and Huse, David A and Bloch, Immanuel and Gross, Christian},
  journal={Science},
  volume={352},
  number={6293},
  pages={1547--1552},
  year={2016},
  publisher={American Association for the Advancement of Science}
}

@article{choi2017observation,
  title={Observation of discrete time-crystalline order in a disordered dipolar many-body system},
  author={Choi, Soonwon and Choi, Joonhee and Landig, Renate and Kucsko, Georg and Zhou, Hengyun and Isoya, Junichi and Jelezko, Fedor and Onoda, Shinobu and Sumiya, Hitoshi and Khemani, Vedika and others},
  journal={Nature},
  volume={543},
  number={7644},
  pages={221--225},
  year={2017},
  publisher={Nature Publishing Group}
}

@article{zhang2017observation,
  title={Observation of a discrete time crystal},
  author={Zhang, Jiehang and Hess, Paul W and Kyprianidis, A and Becker, Petra and Lee, A and Smith, J and Pagano, Gaetano and Potirniche, I-D and Potter, Andrew C and Vishwanath, Ashvin and others},
  journal={Nature},
  volume={543},
  number={7644},
  pages={217--220},
  year={2017},
  publisher={Nature Publishing Group}
}

@article{Ma2019AuthorPhotons,
	title = {A dissipatively stabilized {Mott} insulator of photons},
	volume = {566},
	issn = {1476-4687},
	doi = {10.1038/s41586-019-0897-9},
	number = {7742},
	journal = {Nature},
	author = {Ma, Ruichao and Saxberg, Brendan and Owens, Clai and Leung, Nelson and Lu, Yao and Simon, Jonathan and Schuster, David I.},
	month = feb,
	year = {2019},
	pages = {51--57},
}

@article{Paredes2004TonksGirardeauLattice,
    title = {{Tonks–Girardeau gas of ultracold atoms in an optical lattice}},
    year = {2004},
    journal = {Nature},
    author = {Paredes, Belén and Widera, Artur and Murg, Valentin and Mandel, Olaf and F{\"{o}}lling, Simon and Cirac, Ignacio and Shlyapnikov, Gora V. and H{\"{a}}nsch, Theodor W. and Bloch, Immanuel},
    number = {6989},
    month = {5},
    pages = {277--281},
    volume = {429},
    doi = {10.1038/nature02530},
    issn = {0028-0836}
}

@article{cheuk2015quantum,
  title={Quantum-gas microscope for fermionic atoms},
  author={Cheuk, Lawrence W and Nichols, Matthew A and Okan, Melih and Gersdorf, Thomas and Ramasesh, Vinay V and Bakr, Waseem S and Lompe, Thomas and Zwierlein, Martin W},
  journal={Phys. Rev. Lett.},
  volume={114},
  number={19},
  pages={193001},
  year={2015},
  publisher={APS}
}

@article{browaeys2020many,
  title={Many-body physics with individually controlled Rydberg atoms},
  author={Browaeys, Antoine and Lahaye, Thierry},
  journal={Nature Physics},
  volume={16},
  number={2},
  pages={132--142},
  year={2020},
  publisher={Nature Publishing Group UK London}
}

@Article{Aspuru-Guzik2012quantum,
author={Aspuru-Guzik, Al{\'a}n
and Walther, Philip},
title={Photonic quantum simulators},
journal={Nature Physics},
year={2012},
month={Apr},
day={01},
volume={8},
number={4},
pages={285-291},
issn={1745-2481},
doi={10.1038/nphys2253}
}

@Article{Rigol2008,
author={Rigol, Marcos
and Dunjko, Vanja
and Olshanii, Maxim},
title={Thermalization and its mechanism for generic isolated quantum systems},
journal={Nature},
year={2008},
month={Apr},
day={01},
volume={452},
number={7189},
pages={854-858},
issn={1476-4687},
doi={10.1038/nature06838}
}

@article{
Kaufmanscience2016,
author = {Adam M. Kaufman  and M. Eric Tai  and Alexander Lukin  and Matthew Rispoli  and Robert Schittko  and Philipp M. Preiss  and Markus Greiner },
title = {Quantum thermalization through entanglement in an isolated many-body system},
journal = {Science},
volume = {353},
number = {6301},
pages = {794-800},
year = {2016},
doi = {10.1126/science.aaf6725}}

@Article{Bernien2017,
author={Bernien, Hannes
and Schwartz, Sylvain
and Keesling, Alexander
and Levine, Harry
and Omran, Ahmed
and Pichler, Hannes
and Choi, Soonwon
and Zibrov, Alexander S.
and Endres, Manuel
and Greiner, Markus
and Vuleti{\'{c}}, Vladan
and Lukin, Mikhail D.},
title={Probing many-body dynamics on a 51-atom quantum simulator},
journal={Nature},
year={2017},
month={Nov},
day={01},
volume={551},
number={7682},
pages={579-584},
issn={1476-4687},
doi={10.1038/nature24622}
}

@article{
Schreiber2015,
author = {Michael Schreiber  and Sean S. Hodgman  and Pranjal Bordia  and Henrik P. Lüschen  and Mark H. Fischer  and Ronen Vosk  and Ehud Altman  and Ulrich Schneider  and Immanuel Bloch },
title = {Observation of many-body localization of interacting fermions in a quasirandom optical lattice},
journal = {Science},
volume = {349},
number = {6250},
pages = {842-845},
year = {2015},
doi = {10.1126/science.aaa7432}}

@article{
Joshi2022,
author = {M. K. Joshi  and F. Kranzl  and A. Schuckert  and I. Lovas  and C. Maier  and R. Blatt  and M. Knap  and C. F. Roos },
title = {Observing emergent hydrodynamics in a long-range quantum magnet},
journal = {Science},
volume = {376},
number = {6594},
pages = {720-724},
year = {2022},
doi = {10.1126/science.abk2400}}

@article{guo2021observation,
  title={Observation of energy-resolved many-body localization},
  author={Guo, Qiujiang and Cheng, Chen and Sun, Zheng-Hang and Song, Zixuan and Li, Hekang and Wang, Zhen and Ren, Wenhui and Dong, Hang and Zheng, Dongning and Zhang, Yu-Ran and others},
  journal={Nature Physics},
  volume={17},
  number={2},
  pages={234--239},
  year={2021},
  publisher={Nature Publishing Group UK London}
}

@article{BelyanskyPRR2021,
  title = {Frustration-induced anomalous transport and strong photon decay in waveguide QED},
  author = {Belyansky, Ron and Whitsitt, Seth and Lundgren, Rex and Wang, Yidan and Vrajitoarea, Andrei and Houck, Andrew A. and Gorshkov, Alexey V.},
  journal = {Phys. Rev. Res.},
  volume = {3},
  issue = {3},
  pages = {L032058},
  numpages = {6},
  year = {2021},
  month = {Sep},
  publisher = {American Physical Society},
  doi = {10.1103/PhysRevResearch.3.L032058}
}

@article{VrajitoareaUSC,
  title={Ultrastrong light-matter interaction in a multimode photonic crystal},
  author={Andrei Vrajitoarea and Ron Belyansky and Rex Lundgren and Seth Whitsitt and Alexey V. Gorshkov and Andrew A. Houck},
  year={2024},
  eprint={2209.14972},
  archivePrefix={arXiv},
  primaryClass={quant-ph}
}

@article{Degen_RMP2017,
  title = {Quantum sensing},
  author = {Degen, C. L. and Reinhard, F. and Cappellaro, P.},
  journal = {Rev. Mod. Phys.},
  volume = {89},
  issue = {3},
  pages = {035002},
  numpages = {39},
  year = {2017},
  month = {Jul},
  publisher = {American Physical Society},
  doi = {10.1103/RevModPhys.89.035002}
}

@article{Zoller_PRL2024,
  title = {Essay: Quantum Sensing with Atomic, Molecular, and Optical Platforms for Fundamental Physics},
  author = {Ye, Jun and Zoller, Peter},
  journal = {Phys. Rev. Lett.},
  volume = {132},
  issue = {19},
  pages = {190001},
  numpages = {11},
  year = {2024},
  month = {May},
  publisher = {American Physical Society},
  doi = {10.1103/PhysRevLett.132.190001}
}

@article{cirac2012goals,
  title={Goals and opportunities in quantum simulation},
  author={Cirac, J Ignacio and Zoller, Peter},
  journal={Nature physics},
  volume={8},
  number={4},
  pages={264--266},
  year={2012},
  publisher={Nature Publishing Group UK London}
}

@article{Bharti_RMP2022,
  title = {Noisy intermediate-scale quantum algorithms},
  author = {Bharti, Kishor and Cervera-Lierta, Alba and Kyaw, Thi Ha and Haug, Tobias and Alperin-Lea, Sumner and Anand, Abhinav and Degroote, Matthias and Heimonen, Hermanni and Kottmann, Jakob S. and Menke, Tim and Mok, Wai-Keong and Sim, Sukin and Kwek, Leong-Chuan and Aspuru-Guzik, Al\'an},
  journal = {Rev. Mod. Phys.},
  volume = {94},
  issue = {1},
  pages = {015004},
  numpages = {69},
  year = {2022},
  month = {Feb},
  publisher = {American Physical Society},
  doi = {10.1103/RevModPhys.94.015004}
}

@article{cerezo_NatRev2021,
  title={Variational quantum algorithms},
  author={Cerezo, Marco and Arrasmith, Andrew and Babbush, Ryan and Benjamin, Simon C and Endo, Suguru and Fujii, Keisuke and McClean, Jarrod R and Mitarai, Kosuke and Yuan, Xiao and Cincio, Lukasz and others},
  journal={Nature Reviews Physics},
  volume={3},
  number={9},
  pages={625--644},
  year={2021},
  publisher={Nature Publishing Group UK London}
}

@article{bluvstein2022quantum,
  title={A quantum processor based on coherent transport of entangled atom arrays},
  author={Bluvstein, Dolev and Levine, Harry and Semeghini, Giulia and Wang, Tout T and Ebadi, Sepehr and Kalinowski, Marcin and Keesling, Alexander and Maskara, Nishad and Pichler, Hannes and Greiner, Markus and others},
  journal={Nature},
  volume={604},
  number={7906},
  pages={451--456},
  year={2022},
  publisher={Nature Publishing Group UK London}
}

@article{joshi2023exploring,
  title={Exploring large-scale entanglement in quantum simulation},
  author={Joshi, Manoj K and Kokail, Christian and van Bijnen, Rick and Kranzl, Florian and Zache, Torsten V and Blatt, Rainer and Roos, Christian F and Zoller, Peter},
  journal={Nature},
  volume={624},
  number={7992},
  pages={539--544},
  year={2023},
  publisher={Nature Publishing Group UK London}
}

@article{karamlou2024probing,
  title={Probing entanglement in a 2D hard-core Bose--Hubbard lattice},
  author={Karamlou, Amir H and Rosen, Ilan T and Muschinske, Sarah E and Barrett, Cora N and Di Paolo, Agustin and Ding, Leon and Harrington, Patrick M and Hays, Max and Das, Rabindra and Kim, David K and others},
  journal={Nature},
  volume={629},
  number={8012},
  pages={561--566},
  year={2024},
  publisher={Nature Publishing Group UK London}
}

@Article{google_willow2025,
author={Acharya, Rajeev
and Abanin, Dmitry A.
and Aghababaie-Beni, Laleh
and Aleiner, Igor
and Andersen, Trond I.
and Ansmann, Markus
and Arute, Frank
and Arya, Kunal
and Asfaw, Abraham
and Astrakhantsev, Nikita
and Atalaya, Juan
and Babbush, Ryan
and Bacon, Dave
and Ballard, Brian
and Bardin, Joseph C.
and Bausch, Johannes
and Bengtsson, Andreas
and Bilmes, Alexander
and Blackwell, Sam
and Boixo, Sergio
and Bortoli, Gina
and Bourassa, Alexandre
and Bovaird, Jenna
and Brill, Leon
and Broughton, Michael
and Browne, David A.
and Buchea, Brett
and Buckley, Bob B.
and Buell, David A.
and Burger, Tim
and Burkett, Brian
and Bushnell, Nicholas
and Cabrera, Anthony
and Campero, Juan
and Chang, Hung-Shen
and Chen, Yu
and Chen, Zijun
and Chiaro, Ben
and Chik, Desmond
and Chou, Charina
and Claes, Jahan
and Cleland, Agnetta Y.
and Cogan, Josh
and Collins, Roberto
and Conner, Paul
and Courtney, William
and Crook, Alexander L.
and Curtin, Ben
and Das, Sayan
and Davies, Alex
and De Lorenzo, Laura
and Debroy, Dripto M.
and Demura, Sean
and Devoret, Michel
and Di Paolo, Agustin
and Donohoe, Paul
and Drozdov, Ilya
and Dunsworth, Andrew
and Earle, Clint
and Edlich, Thomas
and Eickbusch, Alec
and Elbag, Aviv Moshe
and Elzouka, Mahmoud
and Erickson, Catherine
and Faoro, Lara
and Farhi, Edward
and Ferreira, Vinicius S.
and Burgos, Leslie Flores
and Forati, Ebrahim
and Fowler, Austin G.
and Foxen, Brooks
and Ganjam, Suhas
and Garcia, Gonzalo
and Gasca, Robert
and Genois, {\'E}lie
and Giang, William
and Gidney, Craig
and Gilboa, Dar
and Gosula, Raja
and Dau, Alejandro Grajales
and Graumann, Dietrich
and Greene, Alex
and Gross, Jonathan A.
and Habegger, Steve
and Hall, John
and Hamilton, Michael C.
and Hansen, Monica
and Harrigan, Matthew P.
and Harrington, Sean D.
and Heras, Francisco J. H.
and Heslin, Stephen
and Heu, Paula
and Higgott, Oscar
and Hill, Gordon
and Hilton, Jeremy
and Holland, George
and Hong, Sabrina
and Huang, Hsin-Yuan
and Huff, Ashley
and Huggins, William J.
and Ioffe, Lev B.
and Isakov, Sergei V.
and Iveland, Justin
and Jeffrey, Evan
and Jiang, Zhang
and Jones, Cody
and Jordan, Stephen
and Joshi, Chaitali
and Juhas, Pavol
and Kafri, Dvir
and Kang, Hui
and Karamlou, Amir H.
and Kechedzhi, Kostyantyn
and Kelly, Julian
and Khaire, Trupti
and Khattar, Tanuj
and Khezri, Mostafa
and Kim, Seon
and Klimov, Paul V.
and Klots, Andrey R.
and Kobrin, Bryce
and Kohli, Pushmeet
and Korotkov, Alexander N.
and Kostritsa, Fedor
and Kothari, Robin
and Kozlovskii, Borislav
and Kreikebaum, John Mark
and Kurilovich, Vladislav D.
and Lacroix, Nathan
and Landhuis, David
and Lange-Dei, Tiano
and Langley, Brandon W.
and Laptev, Pavel
and Lau, Kim-Ming
and Le Guevel, Lo{\"i}ck
and Ledford, Justin
and Lee, Joonho
and Lee, Kenny
and Lensky, Yuri D.
and Leon, Shannon
and Lester, Brian J.
and Li, Wing Yan
and Li, Yin
and Lill, Alexander T.
and Liu, Wayne
and Livingston, William P.
and Locharla, Aditya
and Lucero, Erik
and Lundahl, Daniel
and Lunt, Aaron
and Madhuk, Sid
and Malone, Fionn D.
and Maloney, Ashley
and Mandr{\`a}, Salvatore
and Manyika, James
and Martin, Leigh S.
and Martin, Orion
and Martin, Steven
and Maxfield, Cameron
and McClean, Jarrod R.
and McEwen, Matt
and Meeks, Seneca
and Megrant, Anthony
and Mi, Xiao
and Miao, Kevin C.
and Mieszala, Amanda
and Molavi, Reza
and Molina, Sebastian
and Montazeri, Shirin
and Morvan, Alexis
and Movassagh, Ramis
and Mruczkiewicz, Wojciech
and Naaman, Ofer
and Neeley, Matthew
and Neill, Charles
and Nersisyan, Ani
and Neven, Hartmut
and Newman, Michael
and Ng, Jiun How
and Nguyen, Anthony
and Nguyen, Murray
and Ni, Chia-Hung
and Niu, Murphy Yuezhen
and O'Brien, Thomas E.
and Oliver, William D.
and Opremcak, Alex
and Ottosson, Kristoffer
and Petukhov, Andre
and Pizzuto, Alex
and Platt, John
and Potter, Rebecca
and Pritchard, Orion
and Pryadko, Leonid P.
and Quintana, Chris
and Ramachandran, Ganesh
and Reagor, Matthew J.
and Redding, John
and Rhodes, David M.
and Roberts, Gabrielle
and Rosenberg, Eliott
and Rosenfeld, Emma
and Roushan, Pedram
and Rubin, Nicholas C.
and Saei, Negar
and Sank, Daniel
and Sankaragomathi, Kannan
and Satzinger, Kevin J.
and Schurkus, Henry F.
and Schuster, Christopher
and Senior, Andrew W.
and Shearn, Michael J.
and Shorter, Aaron
and Shutty, Noah
and Shvarts, Vladimir
and Singh, Shraddha
and Sivak, Volodymyr
and Skruzny, Jindra
and Small, Spencer
and Smelyanskiy, Vadim
and Smith, W. Clarke
and Somma, Rolando D.
and Springer, Sofia
and Sterling, George
and Strain, Doug
and Suchard, Jordan
and Szasz, Aaron
and Sztein, Alex
and Thor, Douglas
and Torres, Alfredo
and Torunbalci, M. Mert
and Vaishnav, Abeer
and Vargas, Justin
and Vdovichev, Sergey
and Vidal, Guifre
and Villalonga, Benjamin
and Heidweiller, Catherine Vollgraff
and Waltman, Steven
and Wang, Shannon X.
and Ware, Brayden
and Weber, Kate
and Weidel, Travis
and White, Theodore
and Wong, Kristi
and Woo, Bryan W. K.
and Xing, Cheng
and Yao, Z. Jamie
and Yeh, Ping
and Ying, Bicheng
and Yoo, Juhwan
and Yosri, Noureldin
and Young, Grayson
and Zalcman, Adam
and Zhang, Yaxing
and Zhu, Ningfeng
and Zobrist, Nicholas
and AI, Google Quantum
and {Collaborators}},
title={Quantum error correction below the surface code threshold},
journal={Nature},
year={2025},
month={Feb},
day={01},
volume={638},
number={8052},
pages={920-926},
issn={1476-4687},
doi={10.1038/s41586-024-08449-y}
}

@article{srakaew2023subwavelength,
  title={A subwavelength atomic array switched by a single Rydberg atom},
  author={Srakaew, Kritsana and Weckesser, Pascal and Hollerith, Simon and Wei, David and Adler, Daniel and Bloch, Immanuel and Zeiher, Johannes},
  journal={Nature Physics},
  volume={19},
  number={5},
  pages={714--719},
  year={2023},
  publisher={Nature Publishing Group UK London}
}

@article{bekenstein2020quantum,
  title={Quantum metasurfaces with atom arrays},
  author={Bekenstein, Rivka and Pikovski, Igor and Pichler, Hannes and Shahmoon, Ephraim and Yelin, Susanne F and Lukin, Mikhail D},
  journal={Nature Physics},
  volume={16},
  number={6},
  pages={676--681},
  year={2020},
  publisher={Nature Publishing Group UK London}
}

@article{roberts2024manybody,
  title={Manybody interferometry of quantum fluids},
  author={Roberts, Gabrielle and Vrajitoarea, Andrei and Saxberg, Brendan and Panetta, Margaret G and Simon, Jonathan and Schuster, David I},
  journal={Science Advances},
  volume={10},
  number={29},
  pages={eado1069},
  year={2024},
  publisher={American Association for the Advancement of Science}
}

@article{hauke2016measuring,
  title={Measuring multipartite entanglement through dynamic susceptibilities},
  author={Hauke, Philipp and Heyl, Markus and Tagliacozzo, Luca and Zoller, Peter},
  journal={Nature Physics},
  volume={12},
  number={8},
  pages={778--782},
  year={2016},
  publisher={Nature Publishing Group UK London}
}

@article{andersen2025thermalization,
  title={Thermalization and criticality on an analogue--digital quantum simulator},
  author={Andersen, Trond I and Astrakhantsev, Nikita and Karamlou, Amir H and Berndtsson, Julia and Motruk, Johannes and Szasz, Aaron and Gross, Jonathan A and Schuckert, Alexander and Westerhout, Tom and Zhang, Yaxing and others},
  journal={Nature},
  volume={638},
  number={8049},
  pages={79--85},
  year={2025},
  publisher={Nature Publishing Group UK London}
}

@article{motlagh2024generalized,
  title={Generalized quantum signal processing},
  author={Motlagh, Danial and Wiebe, Nathan},
  journal={PRX Quantum},
  volume={5},
  number={2},
  pages={020368},
  year={2024},
  publisher={APS}
}

@article{pichler2016measurement,
  title={Measurement protocol for the entanglement spectrum of cold atoms},
  author={Pichler, Hannes and Zhu, Guanyu and Seif, Alireza and Zoller, Peter and Hafezi, Mohammad},
  journal={Physical Review X},
  volume={6},
  number={4},
  pages={041033},
  year={2016},
  publisher={APS}
}

@article{cazalilla2011rmp,
  title={One dimensional bosons: From condensed matter systems to ultracold gases},
  author={Cazalilla, Miguel Angel and Citro, Roberta and Giamarchi, Thierry and Orignac, Edmond and Rigol, Marcos},
  journal={Reviews of Modern Physics},
  volume={83},
  number={4},
  pages={1405--1466},
  year={2011},
  publisher={APS}
}

@article{eckardt2017colloquium,
  title={Colloquium: Atomic quantum gases in periodically driven optical lattices},
  author={Eckardt, Andr{\'e}},
  journal={Reviews of Modern Physics},
  volume={89},
  number={1},
  pages={011004},
  year={2017},
  publisher={APS}
}

@article{vrajitoarea_natphys2020,
  title={Quantum control of an oscillator using a stimulated Josephson nonlinearity},
  author={Vrajitoarea, Andrei and Huang, Ziwen and Groszkowski, Peter and Koch, Jens and Houck, Andrew A},
  journal={Nature Physics},
  volume={16},
  number={2},
  pages={211--217},
  year={2020},
  publisher={Nature Publishing Group UK London}
}

@article{vrajitoarea_soundwaves,
  title={Sound Waves in Quantum Fluids of Light},
  author={Vrajitoarea, Andrei and Roberts, Gabrielle and Hazzard, Kaden RA and Simon, Jonathan and Schuster, David I},
  journal={(in preparation)}
}

@book{vrajitoarea2020thesis,
  title={Strongly correlated photonic materials: parametric interactions and ultrastrong coupling in circuit qed},
  author={Vrajitoarea, Marius Andrei},
  year={2020},
  publisher={Princeton University}
}

@article{zhao2025flux,
  title={Flux-Tunable Cavity for Dark Matter Detection},
  author={Zhao, Fang and Li, Ziqian and Dixit, Akash V and Roy, Tanay and Vrajitoarea, Andrei and Banerjee, Riju and Anferov, Alexander and Lee, Kan-Heng and Schuster, David I and Chou, Aaron},
  journal={Physical Review Letters},
  volume={135},
  number={20},
  pages={201002},
  year={2025},
  publisher={APS}
}

@article{haghshenas2025digital,
  title={Digital quantum magnetism at the frontier of classical simulations},
  author={Haghshenas, Reza and Chertkov, Eli and Mills, Michael and Kadow, Wilhelm and Lin, Sheng-Hsuan and Chen, Yi-Hsiang and Cade, Chris and Niesen, Ido and Begu{\v{s}}i{\'c}, Tomislav and Rudolph, Manuel S and others},
  journal={arXiv preprint arXiv:2503.20870},
  year={2025}
}

@article{zeng2025adiabatic,
  title={Adiabatic echo protocols for robust quantum many-body state preparation},
  author={Zeng, Zhongda and Giudici, Giuliano and Senoo, Aruku and Baumg{\~A}{\=I}rtner, Alexander and Kaufman, Adam M and Pichler, Hannes},
  journal={arXiv preprint arXiv:2506.12138},
  year={2025}
}

@article{blais2021circuit,
  title={Circuit quantum electrodynamics},
  author={Blais, Alexandre and Grimsmo, Arne L and Girvin, Steven M and Wallraff, Andreas},
  journal={Reviews of Modern Physics},
  volume={93},
  number={2},
  pages={025005},
  year={2021},
  publisher={APS}
}

@inproceedings{gilyen2019quantum,
  title={Quantum singular value transformation and beyond: exponential improvements for quantum matrix arithmetics},
  author={Gily{\'e}n, Andr{\'a}s and Su, Yuan and Low, Guang Hao and Wiebe, Nathan},
  booktitle={Proceedings of the 51st annual ACM SIGACT symposium on theory of computing},
  pages={193--204},
  year={2019}
}

@article{martyn2021grand,
  title={Grand unification of quantum algorithms},
  author={Martyn, John M and Rossi, Zane M and Tan, Andrew K and Chuang, Isaac L},
  journal={PRX quantum},
  volume={2},
  number={4},
  pages={040203},
  year={2021},
  publisher={APS}
}

@article{garratt2407quantum,
  title={Quantum Algorithm to Prepare Quasi-Stationary States (2024)},
  author={Garratt, SJ and Choi, S},
  journal={arXiv preprint arXiv:2407.07893}
}

\onecolumngrid

\clearpage
\newpage






\onecolumngrid
\newpage
\section*{Supplementary Information}
\appendix
\renewcommand{\appendixname}{Supplement}
\renewcommand{\theequation}{S\arabic{equation}}
\renewcommand{\thefigure}{S\arabic{figure}}
\renewcommand{\figurename}{Supplemental Information Fig.}
\renewcommand{\tablename}{Table}
\setcounter{figure}{0}
\setcounter{table}{0}


\section{System parameters and operating points}
\label{SI:SystemparametersandOperatingpoints}
The device is a one-dimensional array of capacitvely coupled transmon qubits, employed in our previous work in~\cite{AdbPaper, roberts2024manybody}. Each transmon has a tunable resonance frequency in the range $\omega_{01}/2\pi \approx (4-6)\,\text{GHz}$, with on-site interactions set by their anharmonicity $U/2\pi \approx 240\,\text{MHz}$. 
The transmon qubits are frequency-tunable by terminating them with a superconducting quantum interference device (SQUID) loop that is inductively coupled to its individual current bias line used for threading magnetic flux through the loop.
The capacitive coupling between nearest-neighbor transmon sites sets a tunneling rate of $J/2\pi \approx 9 \,\text{MHz}$. For site-resolved readout, each qubit is capacitively coupled to a $\lambda/2$ coplanar waveguide (CPW) resonator. The readout resonators are evenly staggered between $\omega_R/2\pi \approx (7-8)\,\text{GHz}$, with linewidths $\kappa_R/2\pi \approx 100\,\text{kHz}$ and qubit-resonator dispersive couplings in the range $\chi_{qR}/2\pi\approx (0.5-1.6)\,\text{MHz}$. The readout resonators are individually coupled to their own $\lambda/2$ CPW Purcell resonators (filters) that are all coupled to a common CPW feedline. The feedline allows us to send all qubit drive pulses (when the qubits are resolved at different frequencies) and perform frequency multiplexed readout, while the Purcell filters reduces the effect of qubit decay ($T_1$ relaxation) into the feedline environment.

In this paper we use different frequency configurations for the transmon qubits, both ordered and disordered, to operate and probe our synthetic material. The experiments involve tuning the qubits between these frequency configurations, diabatically or adiabatically, as detailed in section SI~\ref{SI:PulseSequences}. The configurations are the following:

\begin{itemize}
    \item \textbf{large disorder staggered configuration}: In this configuration the qubits are staggered in a zig-zag frequency pattern, where neighboring qubits are detuned by $>U$ and next-to-nearest neighbors are detuned by $>2J$. In this stagger we initialize the localized photons and the ancilla state with microwave pulses and perform site-resolved dispersive readout of qubit populations.

    \item \textbf{small disorder staggered configuration}: In this configuration we prepare the localized photons prior to the adiabatic evolution to the ordered lattice. The qubit frequencies are staggered such that the qubits on the left of the ancilla are at higher frequencies than the qubit on the right side. The seven-qubit N00N state is prepared over all sites $Q_0 - Q_6$ and we use $Q_3$ as the ancilla. The five-qubit N00N state is prepared over a subset of the site $Q_2 - Q_6$ and we use $Q_4$ as the ancilla. 

    \item \textbf{inverted small disorder staggered configuration}: This configuration is used in the inverted disorder protocol for preparing the N00N states described in Section~\ref{sec:superpositions}. After preparing the $|\text{solid}\rangle + |\text{fluid}\rangle$ in the ordered lattice, we prepare the N00N state in the disordered (computational) basis by adiabatically ramping the qubits to this inverted configuration where the qubits on the right of the ancilla are at a higher frequency than the qubits on the left side.

    \item \textbf{ordered transistor configuration}: In this configuration the sites are nearly degenerate at the lattice frequency, except for the ancilla qubit detuned that is detuned by $U$. In other words, the $\omega_{01}/2\pi$ frequency of the left and right qubits match the $\omega_{12}/2\pi$ ancilla frequency. In this transistor configuration, the transport of photons is conditioned on the quantum state of the ancilla as explained in Section~\ref{sec:superpositions}.
    
\end{itemize}

Qubit relaxation $T_1$ and decoherence $T^\ast_2$ times are measured as a function of frequency to avoid low-lifetime regions coupled to TLS defects. We use these frequency scans to identify optimal staggered configurations and lattice frequencies. We also monitor $T_1,\,T_2$ over time and find average values $T_1\approx 45$\textmu s, $T_2\approx 1.5$ \textmu s. At the closely degenerate lattice configuration, we expect the qubit $T_2$'s to be enhanced by the flux sweet-spots generated from the avoided crossings of the energy levels. To mitigate decoherence effects through the flux bias lines, where we are sending control signals at different bandwidths, we can consider incorporating on-chip stepped-impedance filters~\cite{zhao2025flux} in future designs.

The list of system parameters and operating points are detailed in Table~\ref{SI:systemparameters}. These parameters were measured using standard experimental techniques in circuit QED, outlined in several modern reviews~\cite{blais2021circuit}.

\section{Ramsey interferometry measurements}
\label{SI:RamseyInterferometryMeasurements}

The many-body Ramsey interferometry measurements are conducted following a procedure similar to our previous work in ref.~\cite{roberts2024manybody}. In essence, it involves performing a Ramsey interference experiment on the ancilla qubit to extract its $|0\rangle \rightarrow |1\rangle$ transition frequency $\omega_{01}$ relative to the frequency tone $\omega_d$ used for driving the ancilla. First, the ancilla qubit is prepared in a superposition $\frac{1}{\sqrt{2}}(|0\rangle + |1\rangle)$ with a microwave $\pi/2$ pulse. After a hold time $\Delta t$, the two superposition states develop a relative phase $\Delta \phi = (\omega_{01} - \omega_d) \Delta t$. Varying the hold time leads to oscillations around the equator of the Bloch sphere, which we can convert to oscillations in qubit population with a second $\pi/2$ pulse. From the fringe oscillations in the measured ancilla occupation $P_{|1\rangle} = \frac{1}{2}(1 + \cos \Delta \phi)$, we extract the qubit frequency $\omega_{01}$.

When applying the quantum-controlled transport protocol for creating N00N states, the ancilla becomes entangled with the other qubits in the lattice during the hold time in the Ramsey sequence. This leads to an enhancement in the accrued relative phase $\Delta \phi = (\omega_{01} + \Delta_{MB} - \omega_d) \Delta t$, registered as an increase in the fringe frequency equal to the energy difference between the qubit frequencies $\Delta_{MB}  = \omega_R - \omega_L$ on the right and left of the ancilla.

The time resolution of the Analog-to-Digital Converter (ADC) used for measuring the ancilla occupation is $1$~ns, which sets a bandwidth limit of $500$~MHz for recording the Ramsey fringe frequency without aliasing. We record a large number of fringe periods within a fraction of the decoherence time budget by setting a frequency difference $(\omega_{01} - \omega_d)/2\pi \simeq 50$~MHz. Detuning the drive frequency by $50$~MHz from the ancilla transition would significantly reduce the amplitude contrast of the Ramsey fringes. Instead, we incorporate a time-dependent virtual phase to our second $\pi/2$ pulse in the Ramsey sequence to virtually change the reference frequency of the qubit drive. As shown in Fig.~\ref{fig:adbNOON}c, we record the Ramsey fringes for $200$~ns in steps of $1$~ns, leading to a frequency resolution of $5$~MHz.


\begin{center}
{\renewcommand{\arraystretch}{1.2}
\begin{table}
\begin{tabular}{|c|c|c|c|c|c|c|c|}

\hline
 \textbf{Qubit} & \textbf{0} & \textbf{1} & \textbf{2} & \textbf{3} & \textbf{4} & \textbf{5} & \textbf{6} \\
\hline
 $U_{\textrm{lattice}}/2\pi$ (MHz) & -241 & -240 & -240 & -231 & -234 & -239 & -240  \\ 
 \hline
 $J_{i, i+1}/2\pi$ (MHz) & 9.062 & 9.032 & 8.842 & 8.936 & 9.023 & 9.040 & --  \\ 
 \hline
 $\omega_{\textrm{large-disorder}}/2\pi - 5310 $ (MHz) & 215 & -265 & 262 & -205 & 309 & -308 & 357 \\ 
 \hline
 \begin{tabular}{@{}c@{}}$\omega_{\textrm{small-disorder}}/2\pi - 5310 $ (MHz) \\ (5 qubit protocol)\end{tabular} & -- & -- & 50 & 100 & 234 & -100 & -50  \\ 
 \hline
 \begin{tabular}{@{}c@{}}$\omega_{\textrm{transistor-config}}/2\pi - 5310 $ (MHz) \\ (5 qubit protocol)\end{tabular} & -- & -- & 0 & 0 & 234 & 0 & 0  \\ 
 \hline
 \begin{tabular}{@{}c@{}}$\omega_{\textrm{inverted-disorder}}/2\pi - 5310 $ (MHz) \\ (5 qubit protocol)\end{tabular} & -- & -- & -50 & -100 & 234 & 100 & 50  \\ 
 \hline
 \begin{tabular}{@{}c@{}}$\omega_{\textrm{small-disorder}}/2\pi - 5310 $ (MHz) \\ (7 qubit protocol)\end{tabular} & 75 & 125 & 100 & 231 & 0 & 0 & 0  \\ 
 \hline
 \begin{tabular}{@{}c@{}}$\omega_{\textrm{transistor-config}}/2\pi - 5310 $ (MHz) \\ (7 qubit protocol)\end{tabular} & 0 & 0 & 0 & 231 & 0 & 0 & 0  \\ 
 \hline
 \begin{tabular}{@{}c@{}}$\omega_{\textrm{transistor-config}}/2\pi - 5310 $ (MHz) \\ (7 qubit protocol)\end{tabular} & 0 & 0 & 0 & 231 & -100 & -125 & -75  \\ 
 \hline
 $T1 (\mu s)$ & 14.6 & 35.5 & 57.7 & 28.4 & 60.3 & 54.7 & 40.0  \\
 \hline
 $T2^* (\mu s)$ & 0.85 & 0.64 & 1.31 & 0.77 & 3.57 & 0.84 & 1.4  \\
 \hline
 single-shot readout fidelity  &0.91  &0.92  &0.93 &0.95  &0.87  &0.92  &0.83  \\
 \hline
 $\omega_{\textrm{read}}/2\pi$ (GHz) &6.197& 6.323& 6.427& 6.556& 6.655& 6.78 & 6.871 \\
 \hline
 $\kappa_{\textrm{read}}/2\pi$ (kHz) & 359 & 553 & 203 & 235 & 292 & 220&
       894 \\
 \hline
 $\chi_{\textrm{rd-qb}}/2\pi$ (MHz) & 0.48 & 1.23 & 0.78 & 1.24 & 0.90 & 1.71 & 0.73   \\ 
 \hline

\hline
\end{tabular}
\caption{
	\textbf{System Parameters}
}
\label{SI:systemparameters}
\end{table}
}
\end{center}

\section{Flux Control and Crosstalk}
\label{SI:FluxControlandCrosstalk}
\subsection{Flux control}

This experiment relies on the precise and fast control of the transmon qubit frequencies that set the energy landscape of the lattice. The on-site qubit frequencies are tuned by applying currents to dedicated flux bias lines inductively coupled to each transmon SQUID. 

There are two types of flux control signals used in this experiment. DC flux bias currents are used for statically tuning the qubit frequencies to a target staggered (disordered) configuration for performing readout and single-qubit microwave control (for injecting photons and changing readout basis). Fast RF flux bias pulses are used for providing additional dynamic frequency tuning (on top of the static disordered configuration) to perform our many-body operations at ns (diabatic control) and \textmu s (adiabatic control) time-scales. The DC and RF flux bias signals are combined with bias tees, anchored to the mixing chamber stage, and routed to each individual flux bias line.

In addition to the on-chip flux bias lines, a solenoid of superconducting niobium-titanium (NbTi) wire is attached (and thermally anchored) to the device package to provide a global magnetic field for arranging the qubit frequencies close to the target staggered configuration. This helps greatly reduce the static DC currents applied to the on-chip bias lines and minimize heating effects.

\begin{figure*}[t] 
	\centering
	\includegraphics[width=0.95\textwidth]{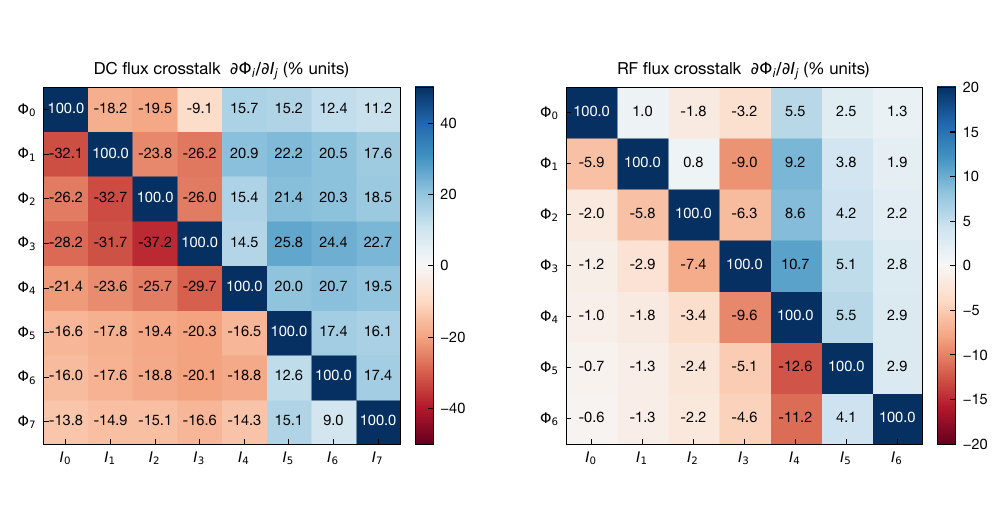}
	\caption{
		\textbf{DC and RF Flux Crosstalk Matrices}
	}
	\label{fig:SI_CTMs}
\end{figure*}

\subsection{DC and RF Crosstalk}

In our device there is non-negligible cross-talk between the flux bias lines.  We need to characterize and calibrate this crosstalk between all qubits and bias lines (at different timescales) to have independent control over each qubit frequency. The amount of flux $\Phi_i$ threading the SQUID loop for a given transmon $Q_i$ is affected not only by its dedicated bias current $I_i$, but also by the currents $I_{j \neq i}$ in all other flux bias lines. 

We assume that any set of flux bias values $\vec{\Phi} = \bold{M} \vec{I}$ is linearly dependent on the applied currents $\vec{I}$ through the crosstalk matrix $\bold{M}$, where the total flux response if given by $d \Phi_i = \sum_{i=1}^N \frac{\partial\Phi_i}{\partial I_j} dI_j$. Each individual matrix element $M_{ij} = \partial\Phi_i/\partial I_j$ corresponds to the mutual inductive coupling between the SQUID loop $i$ and flux control line $j$. We employ the same procedure as in~\cite{AdbPaper, Ma2019AuthorPhotons} for measuring the matrix elements $M_{ij}$. We bias qubit $Q_i$ at a point in its flux dispersion where its frequency $\omega_i^0$ is linearly sensitive to small changes in flux (or current). Using pump-probe qubit spectroscopy, we measure the slope $\partial \omega_i/\partial I_j$ from the variation in qubit frequency when separately varying each bias current $I_j$. The crosstalk matrix element is then calculated by dividing the measured slope by the diagonal variation of the qubit frequency $\omega_i$ with applied flux $\Phi_i$, taking on the general formula $M_{ij} = (\partial \omega_i/\partial I_j)/(\partial \omega_i/\partial \Phi_i)|_{\omega_i^0}$. The qubit frequency to flux conversion is extracted by fitting the measured flux-dependent qubit dispersion $\omega_i = \omega_i (\Phi_i)$ to a Jaynes-Cummings model that takes into account the coupling to individual readout resonators. From the inverted linear dependence on crosstalk $\vec{I} = \bold{M} \vec{\Phi}$, the combination of bias currents $I_j$ needed to independently tune each qubit frequency to any target $\omega_i(\Phi_i)$ is calculated from the eigenvectors of inverted crosstalk matrix $M^{-1}_{ij}$.

The measured crosstalk matrix for static dc flux control is shown in SI Fig.~\ref{fig:SI_CTMs}, where the rows are normalized to the diagonal elements to display the relative magnitude of the off-diagonal crosstalk elements. The fast RF flux crosstalk matrix is also shown in SI Fig.~\ref{fig:SI_CTMs}, measured from the step response of $10$ \textmu s long square pulses with a $1$ ns rise time. We observe minimal drifts ($< 0.2\%$) in the measured DC and RF crosstalk matrices between different cooldowns of the same device and control line configurations.

\subsection{Disorder and Pulse Corrections}

The precision of qubit frequency tuning is limited by the accuracy of our measured crosstalk matrices, flux dispersion $\omega_i(\Phi)$, and cancellation of flux pulse distortions. Using the DC crosstalk matrix, we can reach target frequencies in the staggered configuration to within $\delta \omega_{01}/2\pi \lesssim 10$ MHz. After additional rounds of dispersion corrections $\omega_i(\Phi)$, we reduce the discrepancy to $\delta \omega_{01}/2\pi \lesssim 100$ kHz. Using the RF crosstalk matrix, we can dynamically hit the intended on-site frequencies for our near-degenerate lattice experiments to within $\delta \omega_{01}/2\pi \lesssim 2$ MHz. By measuring the density profiles of our photon fluids at the near-degenerate lattice configuration, and comparing it to theory, we can compensate for additional on-site disorder to further reduce it to $\delta \omega_{01}/2\pi \lesssim 100$ kHz.

In addition to the crosstalk calibrations, the precision in the real-time control of the qubit frequencies (lattice potential) relies on also correcting distortions in the short-time response of the fast RF flux lines. The flux control pulses, used for both diabatic ($\approx 1$ns) and adiabatic ($\approx 100$ns) control, are distorted by the low-pass filters on the RF line. We probe this distortion in spectroscopy by measuring the qubit frequency in response to a flux step pulse and perform a similar kernel correction to the work done in~\cite{AdbPaper, Ma2019AuthorPhotons}.

\section{Pulse Sequences}
\label{SI:PulseSequences}

\begin{figure*}[t] 
	\centering
	\includegraphics[width=0.9\textwidth]{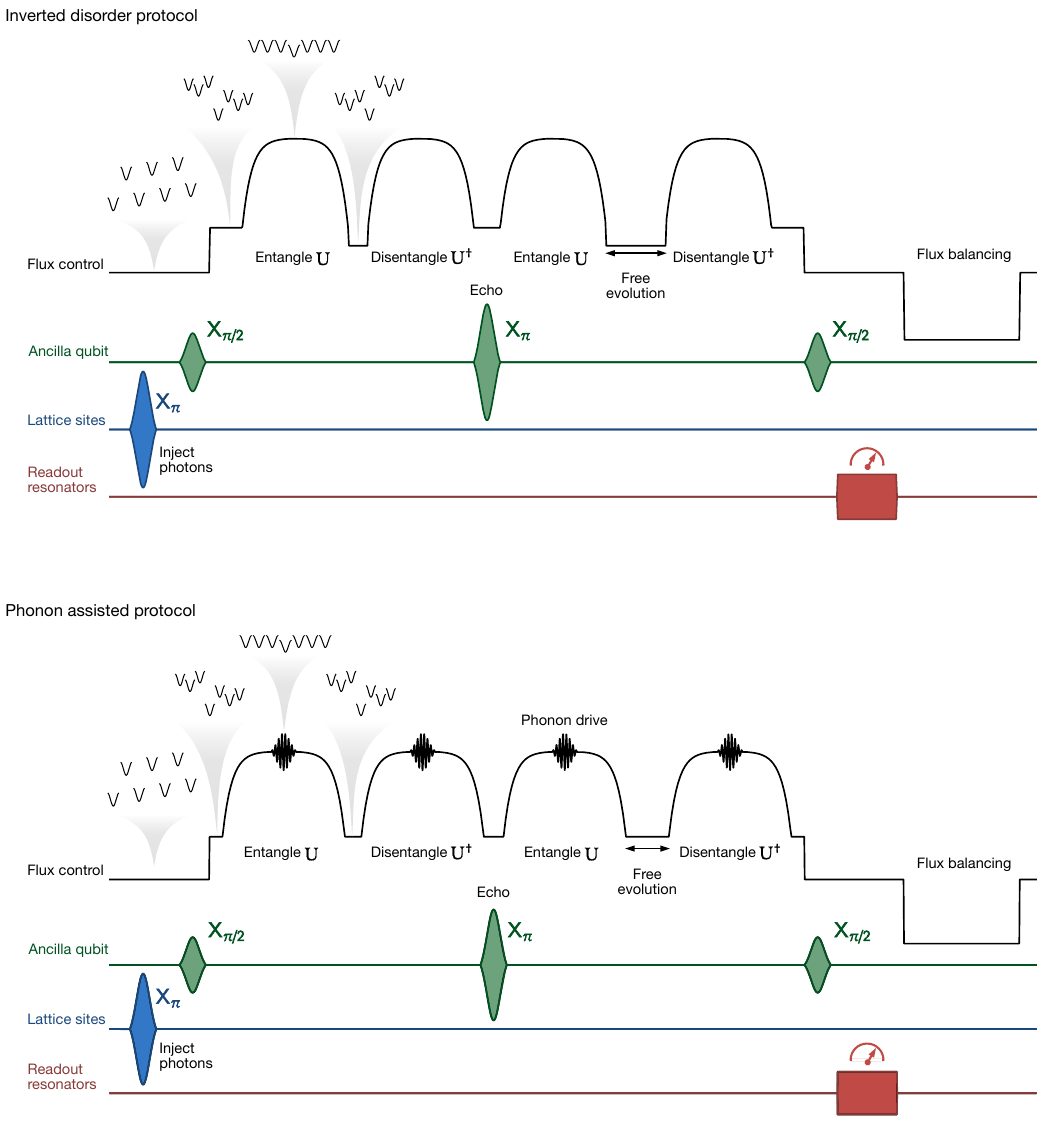}
	\caption{
		\textbf{Pulse sequence} for the drive and flux control pulses used in the two entangling protocols and their corresponding echo sequence
	}
	\label{fig:SI_PulseSequence}
\end{figure*}

The typical pulse sequences used for the conditional transport and many-body interferometry experiments are illustrated in SI Fig.~\ref{fig:SI_PulseSequence}. 
All experiments start with the qubits, in their ground state, DC flux-biased at the large-disordered configuration, where neighboring qubits are detuned by $> U$ (defined in section SI~\ref{SI:SystemparametersandOperatingpoints}). At this staggered configuration we initialize the system with photons by sequentially applying microwave $\pi$-pulses to each frequency-resolved lattice site. We also initialize the ancilla qubit in $|0\rangle$, $|1\rangle$ or $(|0\rangle + |1\rangle)/\sqrt{2}$ using $\pi$ or $\pi/2$ pulses. All microwave qubit pulses have Gaussian envelopes truncated at $\pm 2\sigma$.

Following the injection of photons in the disordered lattice, we rapidly (diabatically) tune the qubits to a smaller staggered configuration where the qubits on the left hand side of the aniclla are at higher frequencies than the ones on the right, and the ancilla qubit is frequency detuned by $U$ from the target lattice frequency. The staggered frequency configurations used for seven and five qubit experiments are detailed in section SI~\ref{SI:SystemparametersandOperatingpoints}. The qubits are then adiabatically ramped to lattice degeneracy in the transistor configuration: the $\omega_{01}/2\pi$ frequency of the left and right qubits matches the $\omega_{12}/2\pi$ frequency of the ancilla. If the ancilla is prepared in the $|0\rangle$, $|1\rangle$ or $(|0\rangle + |1\rangle)/\sqrt{2}$ state, this adiabatic protocol leads to preparing a solid (Mott insulator), fluid (Tonks-Girardeau gas), or coherent superposition of both $|\text{solid}\rangle + |\text{fluid}\rangle$ in the ordered lattice. To probe these states, we freeze tunneling dynamics and onsite occupancy by rapidly (diabatically) tuning the qubits back to the large staggered disorder configuration and apply microwave pulses to each readout resonator to perform site-resolved qubit occupation measurements using heterodyne dispersive readout. For further details regarding the dispersive readout techniques employed for the transmon chain, see the supplemental section of our previous work~\cite{AdbPaper}.

For adiabatically ramping the qubit frequencies we use flux pulses with an exponential shape. This simple approach captures the natural process of the many-gap decreasing as the sites' energies are tuned closer to each other. The total time $t_\text{ramp}$ of these exponential ramps varies with system size. The ramps we use have the functional form $\propto A_\Phi(1 - e^{-t/\tau})$ with the range of timescales $\tau \in \left[\frac{2}{5} t_\text{ramp}, \frac{3}{5} t_\text{ramp}\right]$. We optimize the values for the timescale $\tau$ and total ramp time $t_\text{ramp}$, dependent on the number of photons and lattice size, to satisfy adiabaticity by using the reversibility probe employed in our previous work~\cite{AdbPaper}.

Following the preparation of the $|\text{solid}\rangle + |\text{fluid}\rangle$ state, we can extend the sequence to prepare the N00N state in the disordered (computational) basis. In this work we explore two approaches:
\begin{itemize}
    \item \textbf{inverted disorder protocol} (section~\ref{sec:superpositions}): we adiabatically re-introduce disorder and ramp the qubits to an inverted (\textit{different}) disorder configuration, where the qubits on the right hand side of the ancilla have a higher frequency than the ones on the left (the ancilla qubit remains at the same frequency).
    \item \textbf{phonon assisted protocol} (section~\ref{sec:condphonontransport}): we flux modulate one lattice site to generate particle-hole excitations (phonons) on the fluid component of the superposition to transfer occupation to the fluid state on the opposite edge of the hardcore band (more details in section SI~\ref{SI:PhononTransport}). We re-introduce disorder by ramping to qubits to the \textit{same} staggered configuration we originally started with, where the qubits on the left hand side of the ancilla have a higher frequency.
\end{itemize}

With both of these approaches we prepare a N00N state corresponding to a superposition of photons localized on the left and right hand side of the ancilla. If we wish to probe this entangled state with our many-body Ramsey protocol, we let the system freely evolve at the current frequency stagger, for a hold time which we linearly vary over $\sim 200$ sequences, to accumulate a relative phase that encodes the information about our many-body state. Then we localize this phase information into the ancilla qubit by time-reversing the entanglement pulse sequence, involving two adiabatic ramps and an optional flux modulation tone if employing phonons. We then rapidly tune the qubits to the original large stagger and apply a second $\pi/2$ pulse on the ancilla qubit prior to the readout pulse to convert the oscillations around the equator (from varying the hold time) to fringe oscillations in the qubit occupation. More details are included in section SI~\ref{SI:RamseyInterferometryMeasurements}. This Ramsey sequence is adapted to measure the many-body echo in section~\ref{sec:mbecho} by applying a $\pi$ pulse on the ancilla qubit in the middle of the Ramsey hold time.

The flux bias lines have stray inductances that lead to a very slow ($>1$ms) residual response to flux pulses that induces unwanted qubit frequency drifts. To counteract this effect, at the end of each experimental sequence (after the readout pulse) we apply a flux balancing pulse to cancel the net current flux within one experiment period. 

The experiment pulse sequences are repeated every $500$\textmu s, allowing sufficient idling time for the qubits and readout resonators to decay to their ground state before the start of the next experiment sequence.

\section{Phonon Transport}
\label{SI:PhononTransport}

The key insight in section~\ref{sec:condphonontransport} is the capability of performing quantum-controlled multi-qubit SWAP operations, and leveraging it for preparing N00N states, by exciting phonons in the ordered lattice. This can be intuitively understood from translating the description of the our 1D lattice of strongly interacting photons, described by the Bose-Hubbard Hamiltonian 
\begin{align}
    \mathbf{H}_\mathrm{BH}/\hbar = J \sum_{ \langle i,j \rangle}{a_i^\dagger a_j^{\phantom \dagger}
 } + \frac{U}{2}\sum_i{n_i \left(n_i-1\right)}
+\sum_i {\omega_\mathrm{i} n_i},
\end{align}
in the hard-core limit, to a system of non-interacting fermions~\cite{cazalilla2011rmp, Cazalilla_Tonks_cnt_lat}. To accomplish this, we define operators through a Jordan-Wigner transformation
\begin{align}
    f_i \triangleq \left[\prod_{j<i}(1-2a^\dagger_j a_j^{\phantom \dagger}
)\right]a_i
\end{align}
which obey fermionic anticommutation relations $\{f_i, f_j\} = \delta_{ij}$. The Bose-Hubbard Hamiltonian then becomes a tight-binding model for non-interacting fermions
\begin{align}
    H &= J\sum_{\langle i,j \rangle} f^\dagger_i f_j^{\phantom \dagger}
 + \sum_i \omega_i f^\dagger_i f_i^{\phantom \dagger}
 \\
    &= \sum_k \varepsilon_k f_k^\dagger f_k^{\phantom \dagger}
\end{align}
written in diagonal form using the Fourier transformed fermionic operators $f_k = \frac{1}{\sqrt{N+1}}\sum_j e^{i\frac{\pi k}{N+1}j} f_j$. For a fermionic 1D lattice with $N$ sites and open boundary conditions, the single-particle eigenenergies are $\varepsilon_k = 2J\cos(\frac{\pi k}{N+1})$, where $k\in\{1,N\}$ are the quasi-momenta. 

This description is convenient when describing the many-body states in any $M$-particle ($M<N$) hardcore band, since by simply filling the single-particle fermion states, obeying the Pauli exclusion principle, one produces $M$-particle eigenstates $|\Psi_{k_1,k_2,...,k_M}\rangle  = f^\dagger_{k_1} f^\dagger_{k_2} ... f^\dagger_{k_M} |0\rangle$ with corresponding energies $E_{k_1,k_2,...,k_M} = \sum_{k = k_1}^{k_M} \varepsilon_k$. 
The lowest $|\Psi_\text{min}\rangle$ energy (ground) state simply corresponds to filling up the lowest energy single-particle states, up to the Fermi momentum $|k|<k_F$. The phonons correspond to collective modes (density oscillations) in the fluid that get excited as particle-hole pairs $f^\dagger_{k+q} f_k^{\phantom \dagger}
 |\Psi_\text{min}\rangle$ where one fermion state is taken from the occupied state $k<k_F$ ( resulting in a vacant state $k$, i.e. a hole) and promoted to an occupied state above the Fermi level $k+q>k_F$. The phonon excitation caries a net momentum $q$, and the energy required to excite the mode, on top of the ground state energy, is $\varepsilon_{k+q} - \varepsilon_k$. Due to particle-hole symmetry in our system, the phonon analogy applies also when starting in the highest energy state in the hardcore band by filling up the $M$ highest-energy single-particle fermion states. This becomes relevant as in our experiment we conditionally prepare the highest energy fluid state prior to applying phonon-induced SWAP operations.

For the experiment presented in section~\ref{sec:condphonontransport}, we are looking at $M=2$ particles in $N=5$ sites. The multi-qubit SWAP operation involves transferring photons from the highest frequency transmons on the left side ($Q_2$, $Q_3$) to the lowest frequency one on the right side ($Q_5$, $Q_6$), conditioned on the quantum state of the middle ancilla qubit ($Q_4$). Our adiabatic control maps thesse localized two-photon states in the disordered lattice to the highest and lowest fluids states of the ordered lattice in the transistor configuration ($Q_4$ frequency $\omega_{01}$ detuned by $U$ from the lattice frequency). In terms of fermion operators, the lowest and highest fluid states correspond to $|\Psi_{1,2}\rangle = b^\dagger_{1} b^\dagger_{2} |0\rangle$ and $|\Psi_{4,5}\rangle = b^\dagger_{4} b^\dagger_{5} |0\rangle$, respectively, where the indices denote the quasi-momentum states $\pi k/6$, $k \in \{1,5\}$. The task of swapping photons between sites in the disordered configuration becomes the task of transferring excitations between the extreme fluid states in the hardcore band, which we realize through a two-phonon process $|\Psi_{1,2}\rangle = (b^\dagger_{1} b_4^{\phantom \dagger}
)(b^\dagger_{2} b_5^{\phantom \dagger}
)|\Psi_{4,5}\rangle$ that simultaneously creates two particle-hole pairs between the free fermion states $k=1$ \& $k=4$ and $k=2$ \& $k=5$. 

We drive this two-phonon process by flux-modulating the transmon qubit $Q_3$ with a flat-top gaussian pulse applied through the RF fast-flux line. Since the frequency of the transmon is flux-tunable $\omega_{01} = \omega_{01}(\Phi)$, applying a flux modulation tone $\epsilon_\Phi \cos(\omega_d t)$ on top of the static bias $\Phi_\text{dc}$ leads to a modulation of the transmon frequency $\omega_{01}(\Phi_\text{dc} + \epsilon_\Phi \cos(\omega_d t)) \approx \bar{\omega}_{01} + \epsilon_d \cos(\omega_d t)$, with a modulation amplitude $\epsilon_d = \epsilon_\Phi \,\partial \omega_{01}/\partial \Phi |_{\Phi_b}$.
This flux modulation tone translates to a modulation of the potential at a single site ($Q_3$) $\epsilon_d n_3 \cos(\omega_d t)$, a tool which we can use for resonantly driving excitations in a many-body system~\cite{eckardt2017colloquium}. Matching the flux-drive frequency to the energy required to generate the particle-hole pairs $\omega_d = \varepsilon_4 - \varepsilon_1 = \varepsilon_5 - \varepsilon_2$ leads to coherent Rabi oscillations between the fluid eigenstates $|\Psi_{4,5}\rangle$ and $|\Psi_{1,2}\rangle$ at the edges of the hardcore energy band. This localized lattice perturbation, through the single-site density operator $n_3$, drives resonant transitions between the two many-body states with an effective Rabi coupling rate proportional to $\propto \langle \Psi_{4,5}|n_3|\Psi_{1,2}\rangle J_1(\frac{\epsilon_d}{2\omega_d})$~\cite{vrajitoarea_natphys2020, vrajitoarea2020thesis}, where $J_1$ is the 1st order Bessel function of the first kind.

\begin{figure*}[t] 
	\centering
	\includegraphics[width=1\textwidth]{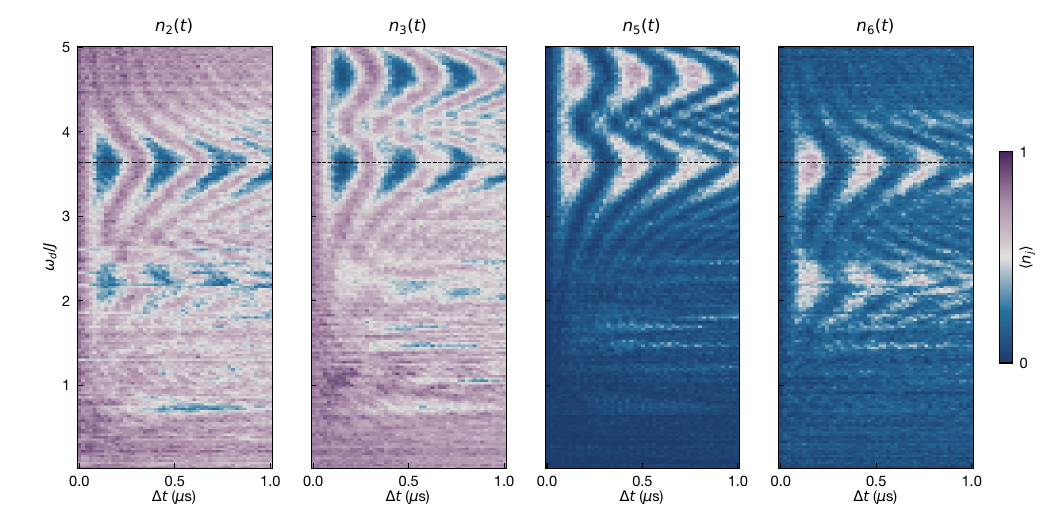}
	\caption{
		\textbf{Phonon-induced SWAP operations} in the disordered (computational) qubit basis, where we are varying the flux modulation tone over a wider range of frequencies.
	}
	\label{fig:SI_MbRabiChevrons}
\end{figure*}

After applying the flux modulation tone, we adiabatically reintroduce disorder with a reversed flux ramp and monitor the qubit population as a function of the frequency and duration of the flux modulation pulse. In SI Fig.~\ref{fig:SI_MbRabiChevrons} we plot the occupation of the transmon qubits $Q_2$, $Q_3$, $Q_5$ and $Q_6$ over a wide range of modulation frequencies (normalized in units of the tunneling frequency $J/2\pi$). In Fig.~\ref{fig:phononSWAP}, the zoomed in chevron oscillations reveal that flux driving at $\omega_d \approx 3.64\,J$ stimulates the desired two-phonon process that exchanges excitations between the fluid states in the ordered lattice (in the transistor configuration). This Rabi oscillations between the few-body fluid states adiabatically translates to multi-qubit SWAP oscillations between the localized photons in the disordered lattice, where we are simultaneously swap photons between the qubits on the left ($Q_2$, $Q_3$) and right ($Q_5$, $Q_6$) side of the ancilla qubit. As a side note, in Fig.~\ref{fig:SI_MbRabiChevrons} we can also see that we are exciting particle-hole pairs between the free fermion states $k=1 \,\&\,k=5$ and $k=2 \,\&\,k=4$ when we drive at $\omega_d\approx 4.7J$ and $\omega_d\approx 2.2J$, respectively, which leads to swapping photons (in the disordered basis) between the qubit pairs $Q_3 \,\&\, Q4$ and $Q_2 \,\&\, Q6$, respectively.

As highlighted in the main text, this phonon mediated SWAP operation is conditioned on the state of the ancilla qubit $Q_4$, as the flux drive generates phonons in the ordered lattice if we prepare a fluid (compressible) state for $Q_4$ in $|1\rangle$, and has no effect when preparing an insulating (incompressible) state for $Q_4$ in $|0\rangle$.

\section{N00N state preparation: adiabatic reversibility}
\label{SI:N00Nfidelity}

\begin{figure*}[t] 
	\centering
	\includegraphics[width=1\textwidth]{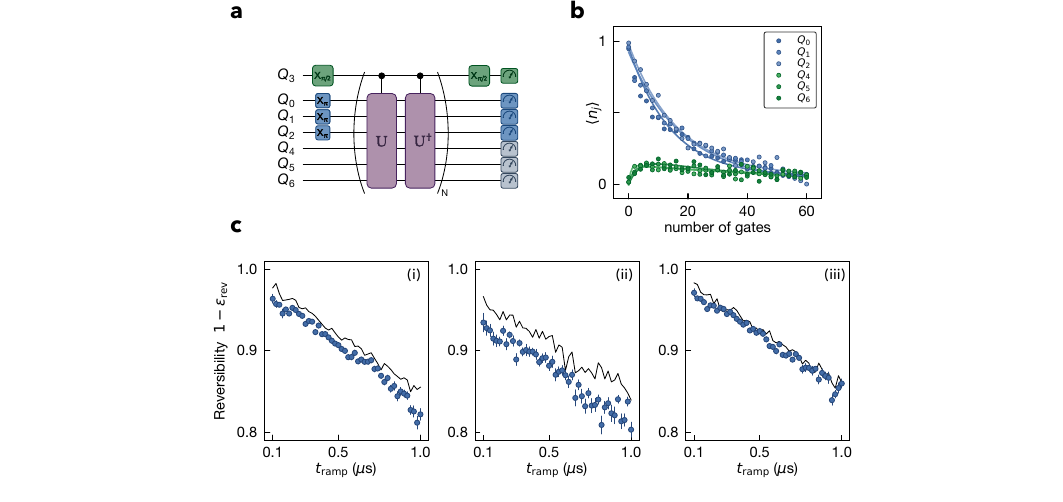}
	\caption{
		\textbf{Fidelity of the adiabatic entangling protocol}. \textbf{a.} We estimate the fidelity of adiabatically preparing the N00N states by applying multiple pairs of entangling ($U$) and disentangling ($U^\dagger$) operations. \textbf{b.} The average error $\epsilon_\mathrm{rev}$ per entangling/disentanling operation is calculated by probing the occupation of the initially excited qubits and fitting the decay with the number of gates. The dataset corresponds to the seven-qubit N00N state. \textbf{c.} The reversibility fidelity $(1-\epsilon_\mathrm{rev})$ is measured as a function of the ramp time $t_\mathrm{ramp}$ for the case of the five-qubit N00N states prepared with the (i) inverted disorder and (ii) phon-assisted protocols, and for the case of the (iii) seven-qubit N00N state prepared with the inverted disorder protocol.
	}
	\label{fig:SI_MbRev}
\end{figure*}

In addition to quantifying the quantum coherence of the prepared N00N state from repeated entangling and disentangling operations in the Ramsey sequenced introduced in section~\ref{sec:mbecho}, we also estimate the errors in the adiabatic evolution of our many-body system by trying to reverse the entangling operation to recover the original product state.  Similarly to the generalized Ramsey sequence, we amplify the error by applying multiple pairs of entangling ($U$) and disentangling ($U^\dagger$) operations after initializing the system with localized photons on one side of the ancilla. We quantify the fidelity from the probability of relocalizing the photons to the initial product state after the last disentangling operation. 
This sequence is highlighted in the diagram in Fig.~\ref{fig:SI_MbRev}a. We measure how the population of the initially excited lattice sites decays with the total number of applied entangling+disentangling operations, as show in Fig.~\ref{fig:SI_MbRev}b for the case of the seven qubit N00N state prepared using the inverted disorder protocol. We extract the average reversibility error $\epsilon_\mathrm{rev}$ by fitting the decay to $\mathcal{F}_\mathrm{adb}(N) = A(1-\epsilon_\mathrm{rev})^{2N}$, where $N$ corresponds to the number of $(U^\dagger \cdot U)$ paired operations and we assume they both have the same average error $\epsilon_\mathrm{rev}$ since the same adiabatic ramp trajectories are employed.

In addition to varying the number of applied entangling/disentangling operations, we also vary the ramp time $t_\mathrm{ramp}$, over an order of magnitude, and measure the fidelity $(1-\epsilon_\mathrm{rev})$ as a function of $t_\mathrm{ramp}$. The results for the five- and seven-qubit N00N states are shown in Fig.~\ref{fig:SI_MbRev}c. For a ramp time of $t_\mathrm{ramp}=240\, \textrm{ns}$, the measured reversibility error rates are $(5.4 \pm 0.2)\%$ and $(4.7 \pm 0.2)\%$ for preparing the five- and seven-qubit N00N states using the inverted disorder protocol, and $(8.1 \pm 0.3)\%$ for preparing the five-qubit N00N state with the phonon assisted protocol. In the case of the phonon assisted protocol, the measured error rates also include the coherent errors in the SWAP operations attributed to imperfect rotation angles.
The measured error rates closely match the decoherence-limited error rates (solid black lines in Fig.~\ref{fig:SI_MbRev}c), which we measure by setting the ancilla qubit to its ground state $|0\rangle$ (decoupling the entanglement dynamics) and running the same sequence of fast flux pulses on all qubits. This suggests that the error rates are primarily limited by $T_1$ decay during the adiabatic ramps. The additional errors likely arise from ancilla decoherence and coherent errors in the phonon SWAP operations.

\end{document}